\documentclass{article}

\usepackage{PRIMEarxiv}

\usepackage[T1]{fontenc}    
\usepackage{hyperref}       
\usepackage{url}            
\usepackage{booktabs}       
\usepackage{amsfonts}       
\usepackage{nicefrac}       
\usepackage{microtype}      
\usepackage{lipsum}
\usepackage{fancyhdr}       
\usepackage{graphicx}       
\usepackage{caption}
\usepackage{comment}
\usepackage{mathtools}
\graphicspath{{media/}}     
\usepackage[dvipsnames]{xcolor}
\usepackage{lineno}
\usepackage[export]{adjustbox} 
\usepackage{authblk}       

\usepackage{fancyhdr}
\newcommand{\headfont}{\bfseries}

\renewcommand{\subsectionmark}[1]{}

\fancypagestyle{plain}{%
\fancyhf{} 

}

\makeatletter
\renewcommand\AB@affilsepx{, \protect\Affilfont}
\makeatother

\title{Anomaly Detection for Automated Data Quality Monitoring in the CMS Detector}
\author[1,5]{Andrew Brinkerhoff}
\author[1]{Chosila Sutantawibul}
\author[6,11]{Robert White}
\author[7]{Caio Daumann}
\author[3,8]{Chad Freer}
\author[2]{Indara Suarez}
\author[2]{Samuel May}
\author[3]{Vivan Nguyen}
\author[4]{Jonathan Guiang}
\author[4]{Bennett Marsh}
\author[5,9]{Darin Acosta}
\author[3]{Alex Aubuchon}
\author[3]{Emanuela~Barberis}
\author[6]{Aaron Bundock}
\author[1]{Evan Collins}
\author[3]{Preston Epps}
\author[7]{Johannes Erdmann}
\author[6]{Henning Flaecher}
\author[1]{Junshen Huang}
\author[2]{Ryan Nie}
\author[6]{Sudarshan Paramesvaran}
\author[5,9]{John Rotter}
\author[2]{Kaitlin Salyer}
\author[1]{Siddhesh Sawant}
\author[10]{Tanvi~Sheokand}
\author[3]{Darien Wood}
\author[ ]{the CMS Muon Detector Collaboration}

\affil[1]{Baylor University}
\affil[2]{Boston University}
\affil[3]{Northeastern University}
\affil[4]{University of California Santa Barbara}
\affil[5]{University of Florida}
\affil[6]{University of Bristol}
\affil[7]{RWTH Aachen University III. Physikalisches Institut A}
\affil[8]{Massachusetts Institute of Technology}
\affil[9]{Rice University}
\affil[10]{Panjab University}
\affil[11]{INFN Sezione di Torino}

\date{January 23, 2025}

\begin{document}

\maketitle

\begin{abstract}
Successful operation of large particle detectors like the Compact Muon Solenoid (CMS) at the CERN Large Hadron Collider requires rapid, in-depth assessment of data quality.
We introduce the ``AutoDQM'' system for Automated Data Quality Monitoring using advanced statistical techniques and unsupervised machine learning.
Anomaly detection algorithms based on the beta-binomial probability function, principal component analysis, and neural network autoencoder image evaluation are tested on the full set of proton-proton collision data collected by CMS in 2022.
AutoDQM identifies anomalous ``bad'' data affected by significant detector malfunction at a rate 4 -- 6 times higher than ``good'' data, demonstrating its effectiveness as a general data quality monitoring tool.
\end{abstract}




\section{Introduction}

The Compact Muon Solenoid (CMS) experiment is a multipurpose particle detector designed to collect large amounts of data from high-energy proton-proton (pp) collisions at the CERN Large Hadron Collider (LHC)~\cite{Chatrchyan:2008zzk,Pettersson:1995yyq}.
The ATLAS and CMS experiments jointly discovered the Higgs boson using LHC collision data collected between 2010 and 2012, and are currently seeking evidence for new physics which could explain dark matter, dark energy, or the matter-antimatter asymmetry of the universe~\cite{Higgs-Discovery_CMS,Higgs-Discovery_ATLAS,Dainese:2019rgk}.

CMS identifies and measures electrons, muons, $\tau$ leptons, photons, and hadrons using a global ``particle-flow'' (PF) algorithm, which combines information from various subdetectors situated concentrically around the proton beam~\cite{CMS:2017yfk}.
A multi-layer silicon tracker traces the helical path of charged particles emerging from the collision point as they are deflected by the CMS solenoid's 3.8~T magnetic field~\cite{CMS:2014pgm}.
Both charged and neutral particles then deposit energy into the electromagnetic calorimeter (ECAL), made from lead tungstate crystals, and the hadron calorimeter (HCAL), made from interleaved brass absorbers and plastic scintillators~\cite{CMS:2020uim,CMS:2018jrd,CMS:2016lmd}.
Beyond the calorimeters, layers of muon tracking detectors, including drift tubes, cathode strip chambers, resistive plate chambers, and gas electron multipliers are embedded in the CMS magnet's flux-return yoke~\cite{CMS:2018rym}.
While the LHC collision rate can exceed 30~MHz, at most around 100~kHz of events pass the initial selection by the CMS Level-1 Trigger (L1T), which uses calorimeter and muon detector inputs to perform preliminary event reconstruction on custom hardware boards in less than $4 \mu$s~\cite{CMS:2020cmk}.
The High-Level Trigger (HLT) then reconstructs events in more detail using CPUs and GPUs, and sends around 1~kHz of collision data to storage on hard disks~\cite{CMS-TRG-19-001}. 

One of the main challenges in operating CMS is to monitor the detector, trigger, and particle reconstruction continuously to ensure that the collected data satisfy the stringent criteria necessary for precise measurements and sensitive searches for new physics phenomena.
Data-taking ``runs'' comprising a few minutes to several hours of collision data are processed in real time to create thousands of histograms which measure various aspects of detector performance.
Trained ``shifters'' monitor these histograms and may intervene if the plots indicate anomalous behavior compared to previous runs.
This data quality monitoring (DQM) is performed for each subdetector system to immediately identify issues ``online'' based on raw data, and also ``offline'' after a few days when full-detector PF reconstruction is available.
Each year a few percent of the total data collected by CMS, corresponding to a dozen or more hours of beamtime, are designated as ``bad'' due to detector or reconstruction issues during data taking.
Without DQM, this fraction would be considerably higher, as problems would go unnoticed for a longer period of time.
Data quality monitoring is thus a time-consuming and labor-intensive but important task, so it is critical to develop robust tools that can help shifters quickly and reliably identify problems in any part of the highly complex CMS detector.

In this paper, we introduce the AutoDQM tool~\footnote{AutoDQM source code publicly available at \url{https://github.com/AutoDQM/AutoDQM}}, a web-based service that employs a generalized approach to automated DQM using statistical techniques and unsupervised machine learning (Section~\ref{sec:autodqm_tool}).
Anomaly detection algorithms based on the beta-binomial probability function, principal component analysis (PCA), and neural network autoencoder (AE) image evaluation are discussed in Sections~\ref{sec:autodqm_stat} and \ref{sec:autodqm_ml}.
AutoDQM performance studies using L1T monitoring plots from the entire 2022 data set are presented in Section~\ref{sec:metastudy}, along with examples of muon detector monitoring using AutoDQM.
These results and plans for future developments are summarized in Section~\ref{sec:summary}.

\section{The AutoDQM tool}
\label{sec:autodqm_tool}

Traditional DQM in CMS is performed using the online and offline DQM web-based GUIs, which contain hundreds of histograms for each CMS subdetector system~\cite{Azzolini:2019asl}.
DQM shifters for each subsystem examine selected histograms in a particular run and look for differences compared to histograms from previous reference runs.
Such anomalies could indicate degraded detector, trigger, or reconstruction performance which would compromise the final physics analysis of the collected data.
Of course, visual comparison of dozens or hundreds of histograms is fatiguing and error-prone.
The AutoDQM anomaly detection tool is a web-based service which evaluates both online and offline DQM histograms and assists shifters in rapid and effective data monitoring.
AutoDQM uses $\chi^2$ and single-bin pull value tests based on the beta-binomial probability function to look for anomalies in a given set of histograms compared to data from one or more previous ``reference'' runs.
PCAs and AEs are also trained on larger sets of reference histograms to generate complex models of good data, which are then used to identify histograms which deviate from the expectation.
AutoDQM graphically represents these statistical and machine learning (ML) tests to highlight anomalous regions within the histograms, allowing shifters and detector experts to quickly identify and locate issues as they arise.

\subsection{Statistical tests}
\label{sec:autodqm_stat}

Most one-dimensional (1D) and two-dimensional (2D) histograms in the CMS DQM GUI have an integer number of entries in each bin.
Depending on the type of histogram and the duration of the data-taking run, a specific histogram may contain millions of entries, or just a few; and these entries may be distributed evenly, or concentrated in a small number of bins.
For a given data histogram, the number of entries $d_i$ in each bin $i$ may be treated as the frequency of a distinct outcome out of $D$ trials, where $D$ is the integral of the histogram.
A reference histogram from a prior run with integral $R$ and $r_i$ entries in each bin can be used to compute the likelihood $\mathcal{L}_i$ to observe $d_i$ in each corresponding bin from the later data run, using the beta-binomial function:

\begin{equation}
  \mathcal{L}_i = f(d_i|D,\alpha,\beta) = {\binom{D}{d_i}} \frac{B(d_i+\alpha,D-d_i+\beta)}{B(\alpha,\beta)}
\label{eq:beta_binom}
\end{equation}
where $B$ is the beta function, $\alpha = \alpha_0 + r_i$, and $\beta = \beta_0 + R - r_i$.
We set $\alpha_0 = \beta_0 = 1$, consistent with uniform priors,
and use the \texttt{betabinom} probability mass function (pmf) implementation in \texttt{SciPy} to compute $\mathcal{L}_i$ for all bins simultaneously using a \texttt{numpy} array representation of the histogram~\cite{scipy_betabinom}.
These $\mathcal{L}_i$ values can be compared to the maximum likelihood $\mathcal{L}_i^{max}$ for each bin, corresponding to $d_i^{max} = D \times r_i / R$ (rounded up or down).
The ratio $\mathcal{L}_i / \mathcal{L}_i^{max}$ gives a relative likelihood $\mathcal{L}_i^{rel}$, which is converted to a pull value $Z_i$ in units of standard deviations using the relation $Z_i^2 = -2~\mathrm{ln}~\mathcal{L}_i^{rel}$.
In order to ensure a minimum ``tolerance'' of $\approx$1\% in the prediction, we scale both $R$ and $r_i$ by a factor of $\tau = 1/\sqrt{1 + (10^{-4}~r_i)^2}$ when computing $\mathcal{L}_i^{rel}$ for each bin, such that $\tau \times r_i \rightarrow 10^4$ as $r_i \rightarrow \infty$, yielding a minimum uncertainty of about $1 / \sqrt{10^4} = 1\%$.
When comparing a data run to multiple reference runs, $Z_i$ is derived using the average of the $\mathcal{L}_i^{rel}$ values, which are computed separately with respect to each reference run.
With this approach, if the observed data matches the expectation from at least one of the reference histograms well, the pull values will not be very large.
For example, if seven reference runs each give $\mathcal{L}_i^{rel} \approx 0$ ($Z_i \approx \infty$) for a given bin, but one reference run gives $\mathcal{L}_i^{rel} = 0.33$ ($Z_i \approx 1.5$), the final $Z_i = \sqrt{-2~\mathrm{ln}~(0.33 / 8)} \approx 2.5$, which is larger than 1.5 but much smaller than $\infty$.
This allows the statistical test to account for systematic variations in histogram shapes due to changing collision conditions between runs.

\begin{figure}[htbp]
\begin{center}
\includegraphics[width=0.49\textwidth]{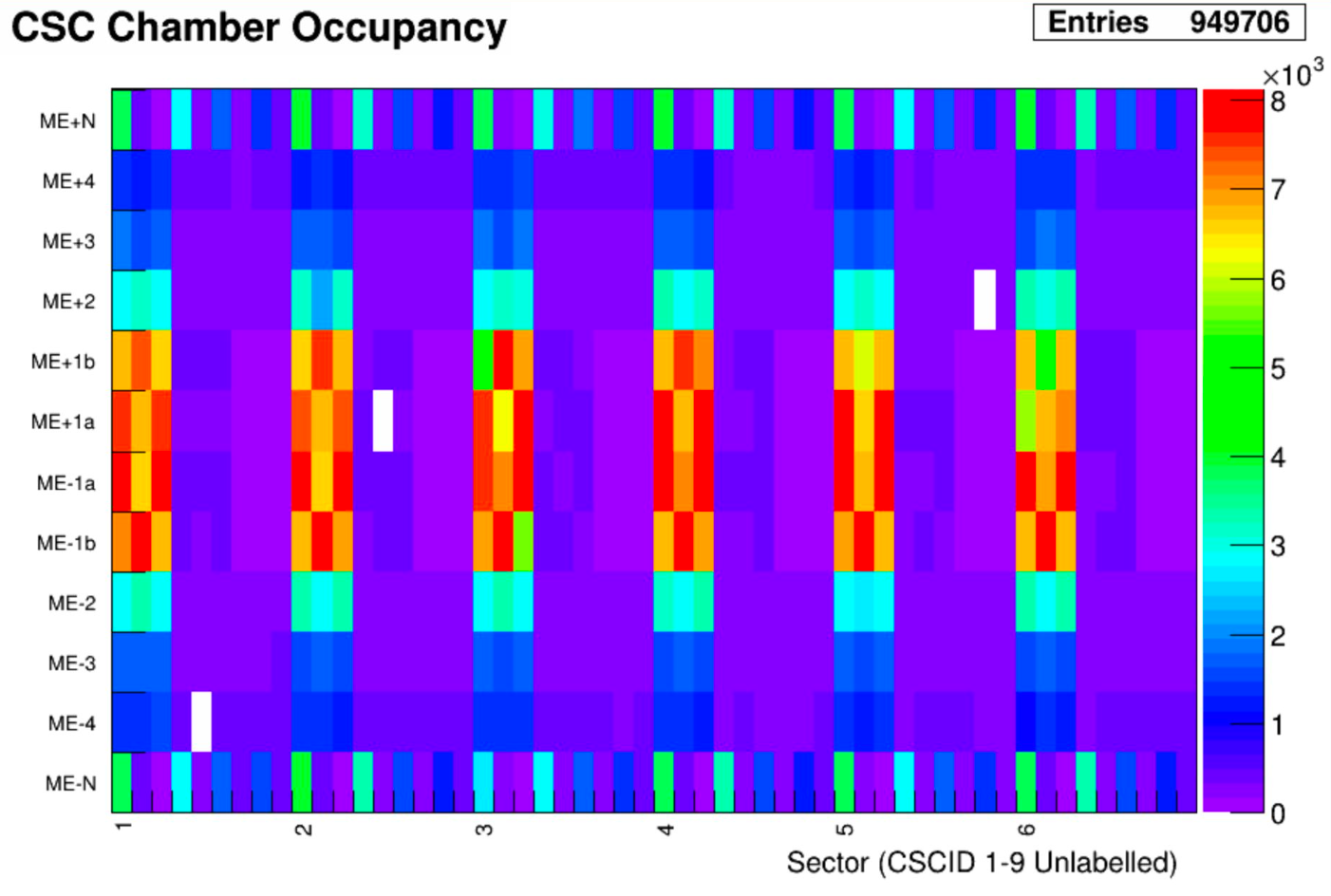}
\includegraphics[width=0.49\textwidth]{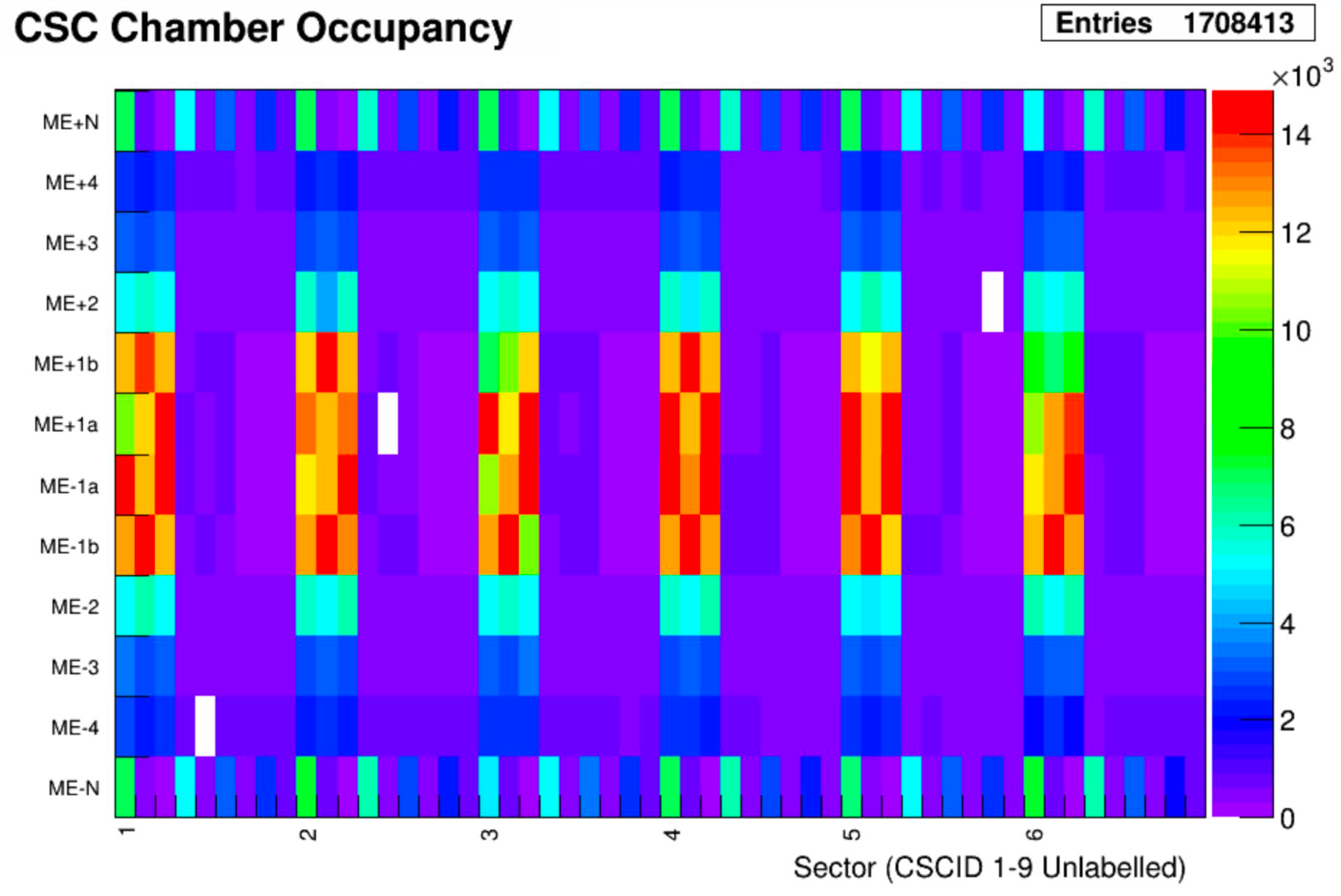}
\includegraphics[width=0.49\textwidth]{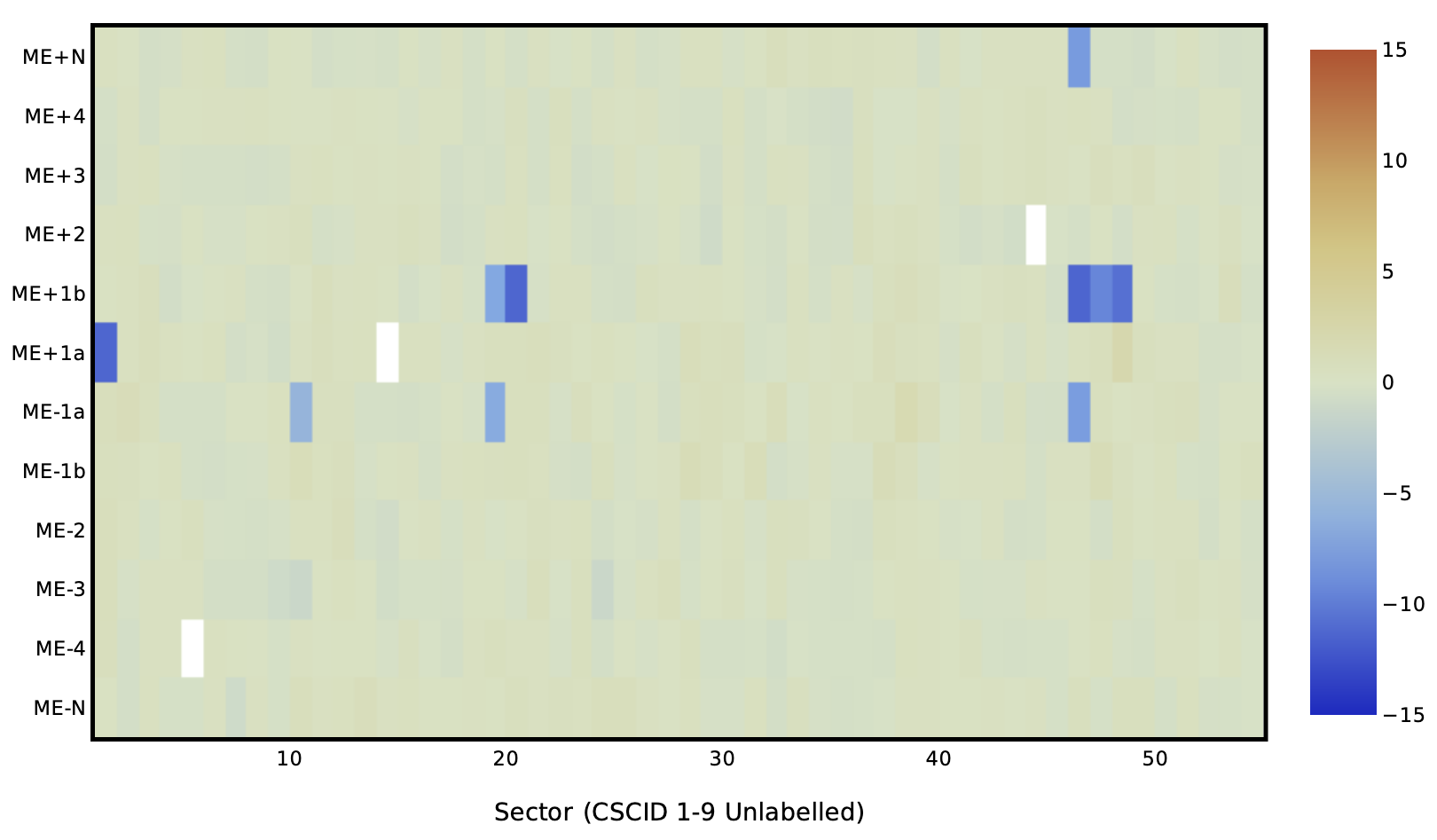}
\end{center}
\caption{A 2D muon track ``stub'' occupancy histogram for cathode strip chambers (CSCs) in reference run 356937 (left), data run 357001 (right), and the AutoDQM heat map showing regions of statistically significant deficits in blue when comparing the data run to 8 prior ``good'' reference runs (bottom).
These deficits in run 357001 are almost invisible in the original DQM GUI histogram. }
\label{fig:CSC_TP_2D}
\end{figure}

\begin{figure}[htbp]
\begin{center}
\includegraphics[width=0.42\textwidth, valign=t]{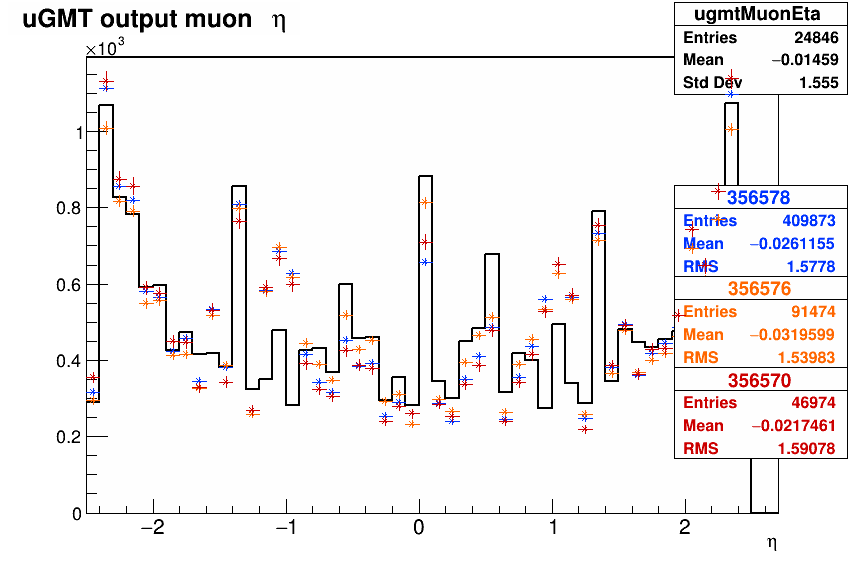}
\includegraphics[width=0.56\textwidth, valign=t]{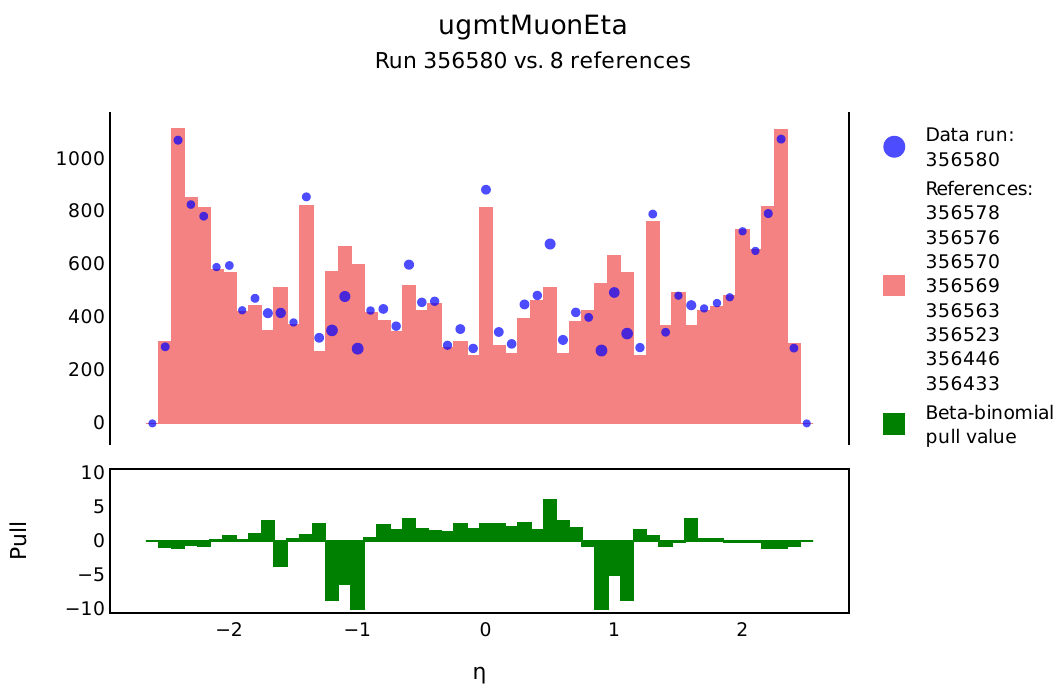}
\includegraphics[width=0.42\textwidth, valign=t]{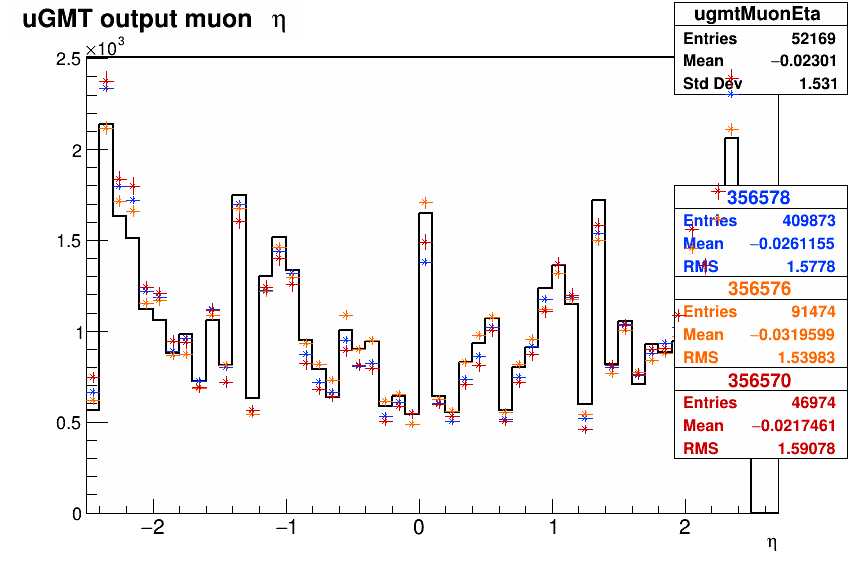}
\includegraphics[width=0.56\textwidth, valign=t]{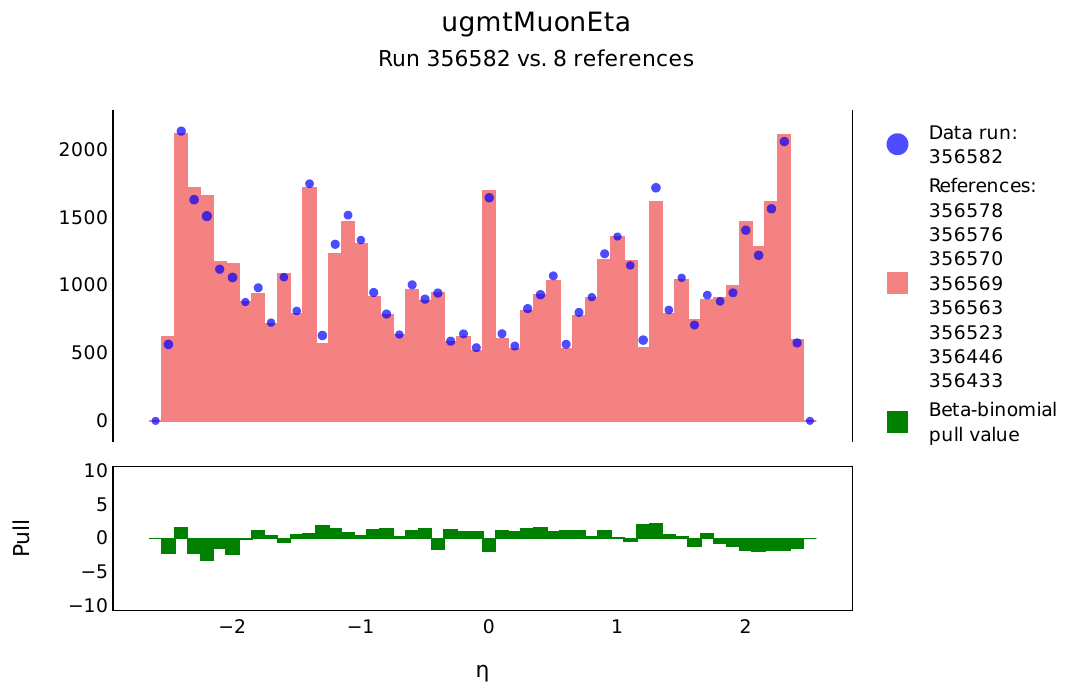}
\end{center}
\caption{The standard DQM histogram for the pseudorapidity distribution $\eta$ of reconstructed muon tracks in the L1T from data run 356580 in black, overlaid with normalized distributions from 4 previous reference runs (upper left).
AutoDQM flags the data as anomalous compared to 8 previous reference runs, and makes the local deficit more visible with the beta-binomial pull value histogram in green (upper right).
The corresponding plots for run 356582, where the muon detector issue was resolved, are shown below. }
\label{fig:OMTF_1D}
\end{figure}

\begin{figure}[htbp]
\begin{center}
\includegraphics[width=0.49\textwidth]{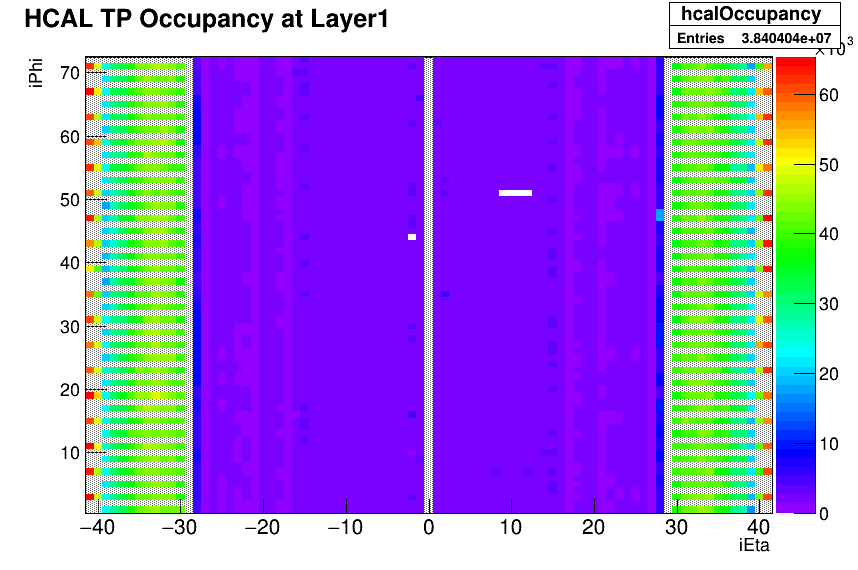}
\includegraphics[width=0.49\textwidth]{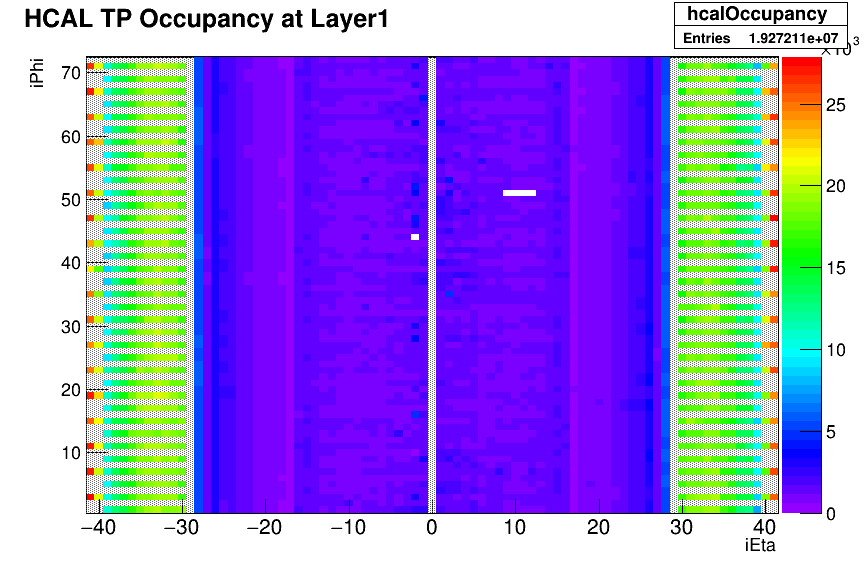}
\includegraphics[width=0.49\textwidth]{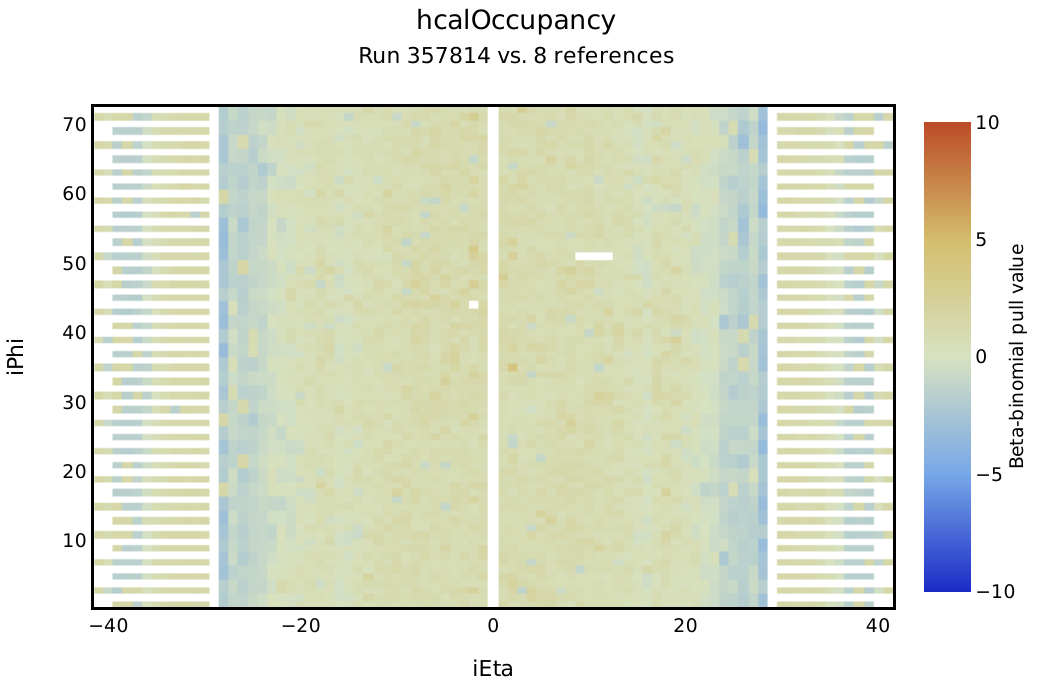}
\includegraphics[width=0.49\textwidth]{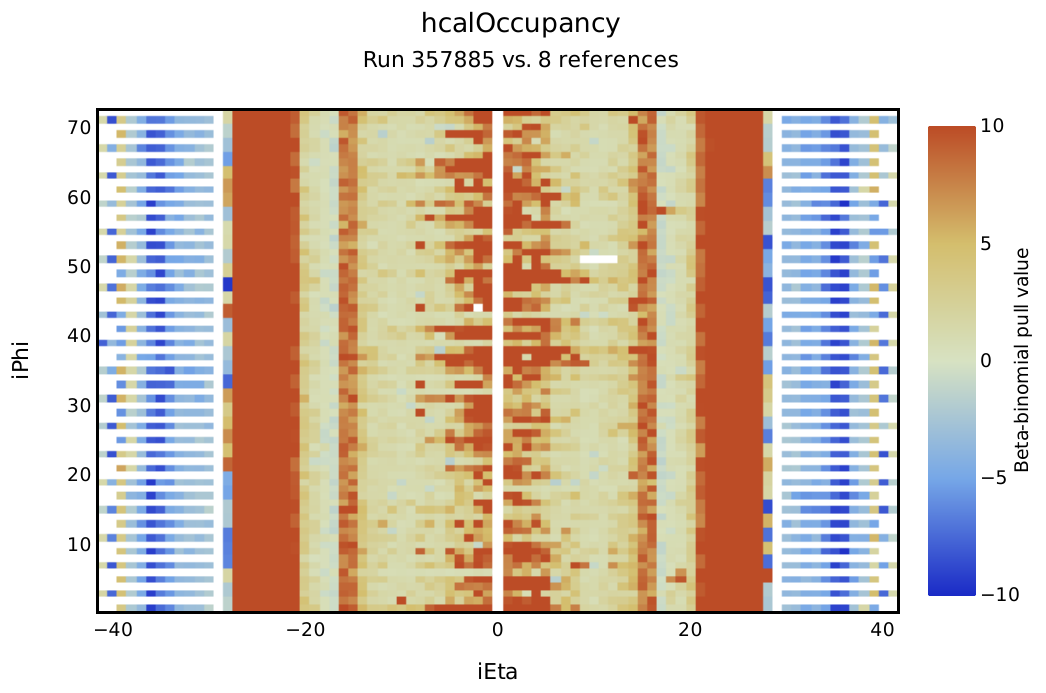}
\end{center}
\caption{The 2D $\phi$ vs.\ $\eta$ geometrical distributions of energy deposits in the HCAL detector transmitted to the L1T, from data runs 357814 (upper left) and 357885 (upper right).
There is very little visible difference between these DQM histograms, which are monitored daily by L1T experts.
The AutoDQM beta-binomial pull value ``heat map'' indicates no anomalous behavior for run 357814 when comparing to 8 previous reference runs (lower left), but correctly flags run 357885, which had an issue with HCAL timing (lower right).}
\label{fig:HCAL_2D}
\end{figure}

In the AutoDQM GUI, the pull values for each bin in a 2D histogram are displayed as a heat map, allowing shifters to quickly identify the location of any significant excess (red) or deficit (blue), as shown in Figs.~\ref{fig:CSC_TP_2D} and \ref{fig:HCAL_2D}.
For 1D histograms, the data distribution in blue overlays a per-bin probability-weighted average of the reference distributions in red, with the pull values in a separate panel in green (Fig.~\ref{fig:OMTF_1D}).
Only histograms identified as anomalous are displayed immediately to shifters, focusing their attention on confirmed discrepancies.
Shifters can view the full set of histograms by clicking a button in the GUI (Fig.~\ref{fig:CSC_webpage}).

The first statistical anomaly metric used by AutoDQM is $\chi^2 = \sum Z_i^2 / nBins$.
The second anomaly metric is the modified maximum pull magnitude $Z'_{max}$ out of all the bins, where the smallest single-bin relative likelihood $\mathcal{L}_{i~min}^{rel}$ is first adjusted for the look-elsewhere effect: $\mathcal{L}_{i~min}^{rel~\prime} = 1 - (1 - \mathcal{L}_{i~min}^{rel})^{nBins}$.
The $\chi^2$ and $Z'_{max}$ thresholds to flag histograms as anomalous can be configured independently for each type of histogram.


\subsection{Machine learning for DQM anomaly detection}
\label{sec:autodqm_ml}

Machine learning algorithms can detect anomalies without the need for specific reference data, while also accounting for expected systematic variations in the histogram shapes.
AutoDQM uses two different unsupervised ML algorithms, based on Principal Component Analysis (PCAs) and autoencoders (AEs).
This is preferred over the supervised approach, in which algorithms are trained with explicitly labeled ``good'' and ``bad'' data, for two reasons.
First, bad data are rare -- many detector subsystems do not have enough bad data to effectively frame this as a supervised problem.
Second, past problems resulting in bad data may not be representative of future issues.
The unsupervised approach only requires a sufficient quantity of good data to train the algorithm, and is agnostic to the particular type of anomaly which could indicate bad data.

More formally, our unsupervised approach to DQM anomaly detection can be framed as follows: given a collection of histograms from good runs, we seek to learn a transformation $\mathcal T$ from the input space (i.e. the entries in each bin) into a lower-dimensional ``latent space'', such that the latent space can be used to approximately reconstruct the original histogram.
The form of the transformation $\mathcal T$ and the cost function used to optimize its parameters differ according to the ML algorithm used, as described for PCAs and AEs in sections~\ref{sec:autodqm_pca} and \ref{sec:autodqm_ae}, respectively.

Once a transformation $\mathcal T$ has been learned, the anomaly score for data histogram $d$ (normalized to area 1) can be calculated as the sum of the squares of the errors (SSE) between the original and reconstructed histograms:
\begin{equation}
\label{eq:sse}
   \mathrm{SSE} = \sum_{i=1}^{nBins} (d'_i - d_i)^2,
\end{equation}
where $d_i$ are the contents of bin $i$ of the normalized data histogram, and $d'_i$ are the bin contents obtained after applying the transformation $\mathcal T$ and subsequently its inverse:
$d' = \mathcal T^{-1} (\mathcal T d)$.
Histograms that match the training data have a self-similar reconstruction under transformation $\mathcal T^{-1} \mathcal T$, leading to low SSE scores, while those with significant deviations compared to the training set do not.
In this way, bad runs which behave differently than the good runs used in training can be identified by DQM histograms with high SSE scores, regardless of the nature of the anomaly.

One shortcoming of the normalized SSE metric is its anti-correlation to the number of entries in a histogram, such that histograms in shorter runs consistently get higher anomaly scores (see Appendix~\ref{sec:appendix}).
To address this, we instead use a $\chi^2$ metric, which is intrinsically less sensitive to statistical fluctuations than the SSE.
We use the beta-binomial probability function (Eq.~\ref{eq:beta_binom}) with the original data histogram $d$ (integral $D$, not normalized to 1), and take $d'$ (integral $D'$, again not normalized) scaled by 100 as the ``reference'' histogram; so mathematically:
\begin{equation}
  \alpha = \alpha_0 + d'_i \times 100~\mathrm{and}~\beta = \beta_0 + D' \times 100,~\mathrm{with}~\alpha_0 = \beta_0 = 1
\label{eq:beta_binom_PCA}
\end{equation}
The factor of 100 suppresses any statistical uncertainty in the transformed bin contents $d'_i$, leaving only the uncertainty in the original $d_i$. 
However, this $\chi^2$ score is consistently $low$ for histograms with fewer entries, so a modified version is used, scaling by the number of entries $D$:
\begin{equation}
  \chi^{2~\prime}  = \chi^{2}~/~D^{1/3}.
\label{eq:mod_chi2}
\end{equation}
This scaling mitigates the $\chi^2$ score dependence on $D$, as shown in Appendix~\ref{sec:appendix}.

\subsubsection{Principal component analysis}
\label{sec:autodqm_pca}
Principal component analysis is an unsupervised ML approach frequently used in the context of dimensionality reduction~\cite{Pearson01111901}.
AutoDQM uses the PCA implementation from \texttt{scikit-learn}~\cite{scikit-learn}.
The PCA takes a 1D or 2D histogram from a given run as input, and transforms it into 2 key components extracted from the training set of 216 good runs.
2D histograms are first flattened into 1D for both training and evaluation.
We found that merging low-occupancy bins improves the PCA reconstruction by reducing the impact of statistical fluctuations.
Merging proceeds iteratively until each bin contains at least 0.33\% of the histogram entries, averaging the bin occupancy over the full training dataset.
The reconstruction is produced from the latent space via re-transformation, with negative bins set to zero using a rectifier function as introduced in Ref.~\cite{pmlr-v15-glorot11a}.
This avoids non-physical reconstructed histograms, and is applied in the last stage to avoid biasing the weights in the PCA re-transformation workflow.

The PCA reconstruction ought to closely agree with the input histogram for good histograms, but anomalous features will not be identified as principal components in the initial transformation, and thus will not appear in the same way in the reconstruction.
By comparing the input and reconstructed histograms, the PCA can flag anomalous histograms based on high $\chi^{2~\prime}$ scores (Eq.~\ref{eq:mod_chi2}), typically $> 0.4$, as shown in Fig.~\ref{fig:algo_recon_pca_ae}.


\subsubsection{Autoencoder}
\label{sec:autodqm_ae}
\begin{figure}[htbp]
\begin{center}
\includegraphics[width=0.45\textwidth]{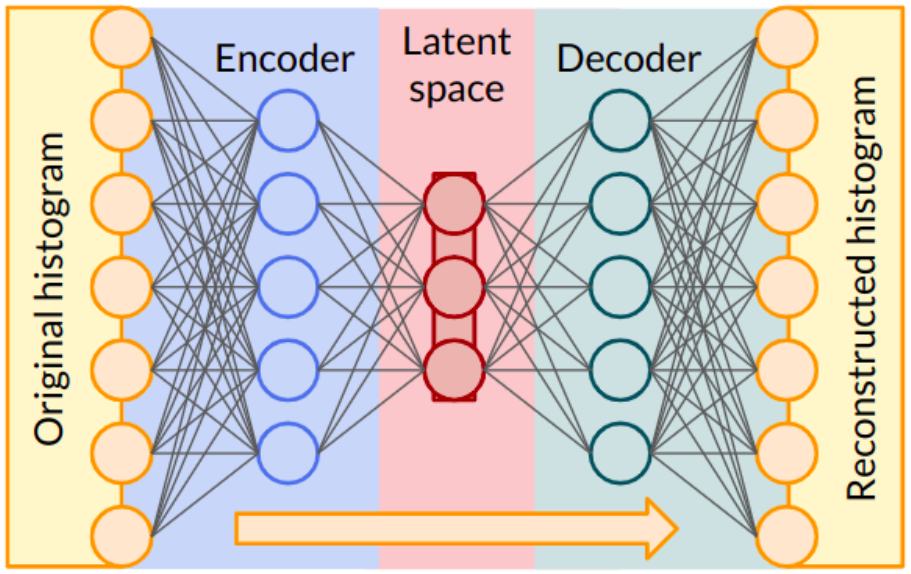}
\end{center}
\caption{A diagram of the autoencoder workflow for a given input histogram, which is compressed in the encoder until a latent-space representation of the input is learned, then is transformed in the decoder to give a reconstructed version of the input histogram for comparison.}
\label{fig:autoencoder_flow}
\end{figure}

An autoencoder is a type of NN that can learn an efficient representation of high-dimensional data. 
The AE has three components: (1) the encoder, which reduces the dimensionality of the input data, (2) the latent space (bottleneck), which represents the input data using even fewer dimensions, and (3) the decoder, which transforms the latent representation back to the original input space.
A rectifier function is also applied to the last layer of decoding, to remove negative bin counts. The AE architecture is visualized in Fig.~\ref{fig:autoencoder_flow}.

\begin{figure}[htbp]
\begin{center}
\includegraphics[width=0.49\textwidth]{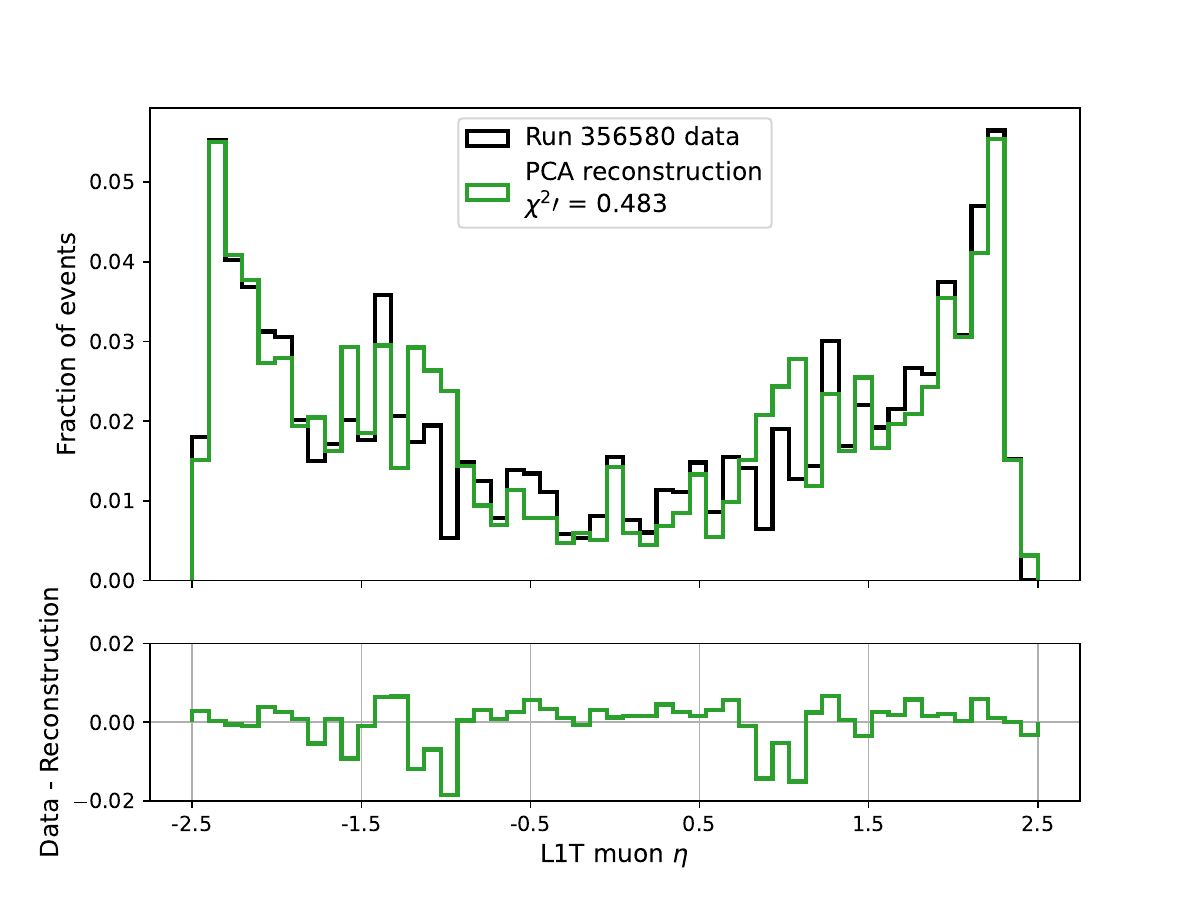}
\includegraphics[width=0.49\textwidth]{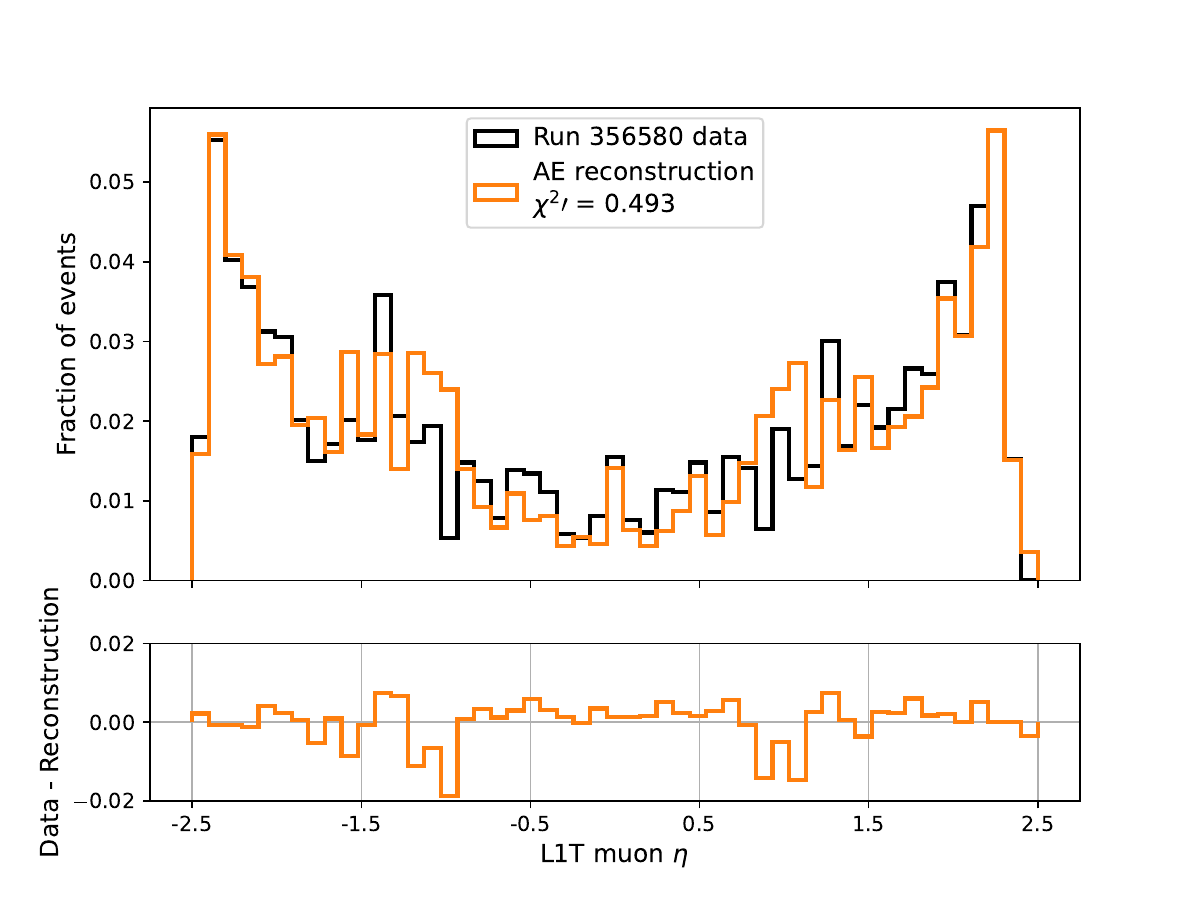}
\includegraphics[width=0.49\textwidth]{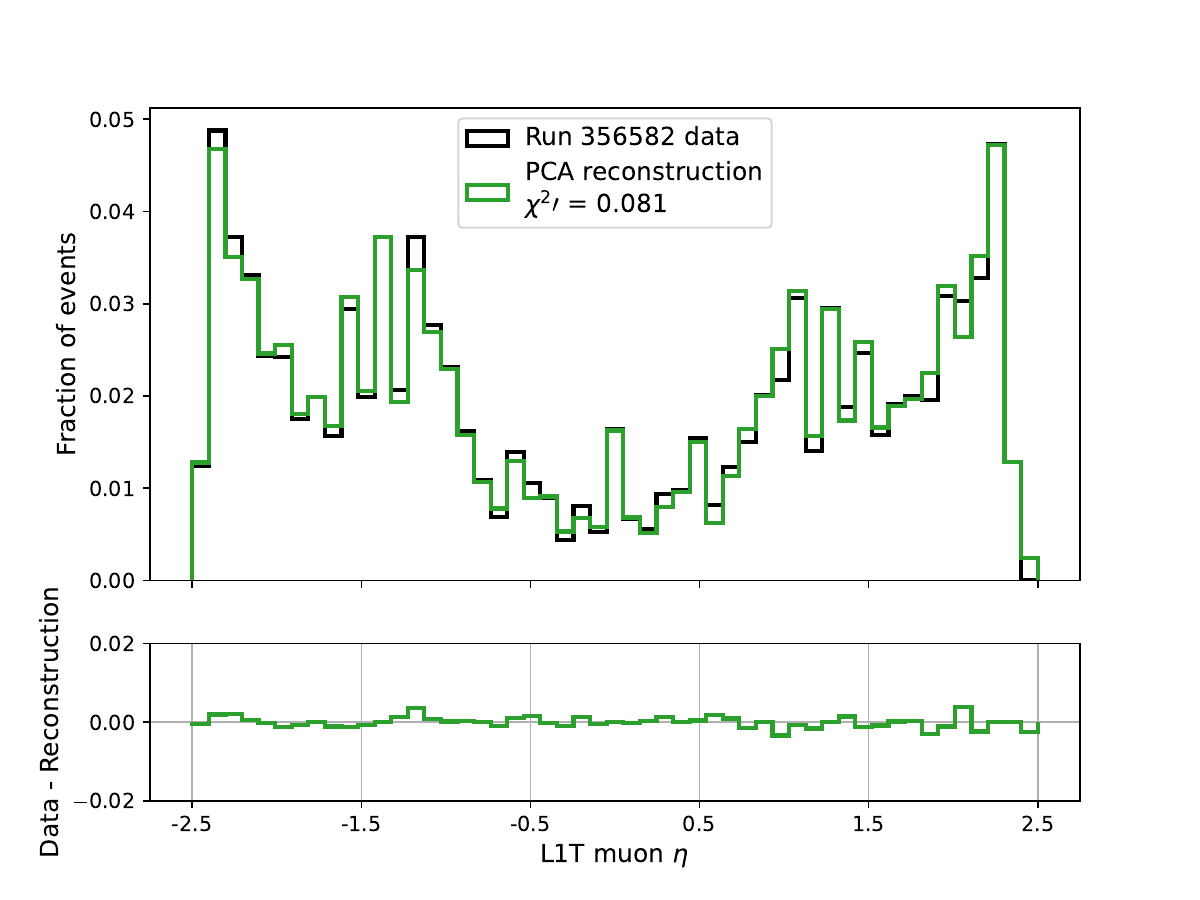}
\includegraphics[width=0.49\textwidth]{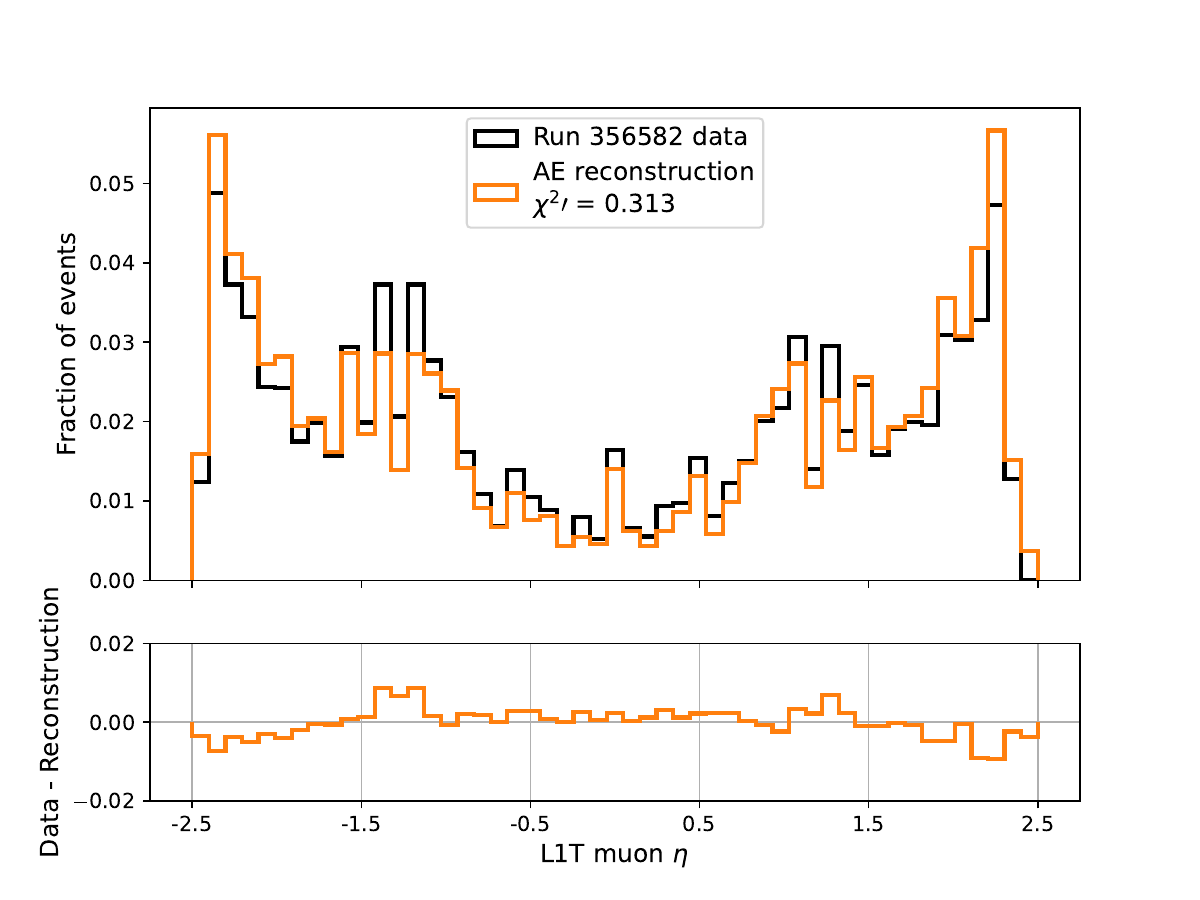}
\end{center}
\caption{Normalized reconstructions of the $\eta$ distribution of muon tracks from runs 356580 (top) and 356582 (bottom), using the PCA (left) and autoencoder (right).
A deficit of tracks in $0.9 < |\eta| < 1.2$ in run 356580 is indicated by the PCA $\chi^{2~\prime}$ score over 0.4, and the data-reconstruction difference in the lower panel for both the PCA and autoencoder.
}
\label{fig:algo_recon_pca_ae}
\end{figure}

The AEs use 1D convolution layers in the encoder and transposed 1D convolution layers in the decoder.
The 2D-to-1D histogram flattening and low-occupancy bin merging used for PCAs is applied for AEs as well, so that 1D and 2D histograms use the same encoder and decoder layers.
An ideal AE will be sensitive enough that it produces a good reconstruction of input data, but insensitive enough to not memorize or overfit during training.
AutoDQM uses the AE implementation from \texttt{tensorflow}~\cite{tensorflow2015-whitepaper}, with 50 nodes, 12 filters, 2 hidden layers, and a learning rate of 0.001.
The 1D input data histogram is overlaid with the reconstructed histogram from the AE, and the $\chi^{2~\prime}$ value is computed,
allowing shifters to see the size and location of any anomalies (Fig.~\ref{fig:algo_recon_pca_ae}).

\section{Performance evaluation}
\label{sec:metastudy}

\subsection{Assessment strategy}
\label{sec:metastudy_strategy}

When developing anomaly detection tools, it is often difficult to rigorously measure their performance identifying true anomalies in data.
An unbiased assessment using real anomalies in real data requires an independent measure of ``anomalousness'' on the same set of data.
Ideally, this independent measure would also focus on \textit{important} anomalies, as many types of data variation have no bearing on whether the data is ``good'' -- in the case of CMS, usable for later physics analysis.
Previous anomaly detection studies for CMS DQM have either used histograms which were individually labeled as bad by visual inspection, or have generated artificial anomalies to mimic problematic detector behavior~\cite{Borisyak_2017,pol2018detector,thecmsecalcollaboration2023autoencoderbased,Asres_2023}.

To measure the AutoDQM performance, we use a full year's worth of data collection runs which were labeled as good or bad by the CMS Physics Performance and Datasets (PPD) group.
The PPD team synthesizes information about each run from detector subsystem experts and analyzers who study reconstructed hadron jets, $\tau$ leptons, photons, electrons, and muons.
PPD then decides whether any detector or reconstruction issues observed are serious enough to exclude all or part of the data in a run from CMS physics analyses.
The final determination is made without reference to the AutoDQM tool, and frequently uses information which is not available in the DQM GUI at all.
Thus the PPD evaluation is both independent of AutoDQM, and reflects the seriousness of anomalous behavior in the CMS detector.

Our assessment data set includes 265 good and 43 bad runs collected in 2022, representing an integrated luminosity of 36~fb$^{-1}$.
Selected runs are required to have lasted at least 5 minutes and contain at least 3~pb$^{-1}$ of collision data, to ensure a sufficient number of entries in the DQM histograms.
Good runs must have at least 90\% of their data labeled good by PPD; bad runs must be over 50\% bad, or contain over 40~pb$^{-1}$ of bad data.
For each run, we examine 62 histograms from the L1T online DQM, covering inputs from the ECAL, HCAL, and muon chambers.
Typical histograms produced for L1T reconstructed hadron jets, electrons and photons, $\tau$ leptons, and muons include 1D and 2D distributions of their $\eta$ and/or $\phi$ location, transverse momentum ($p_{T}$), and isolation- and identification-related quantities.
Because the L1T does not receive any information from the silicon tracker, we exclude runs that PPD labeled bad due to tracker issues.

\subsection{Assessment metrics}
\label{sec:metastudy_metrics}

For each histogram in each run, we compute the $\chi^2$ and $Z'_{max}$ anomaly score values from the beta-binomial statistical test (Eq.~\ref{eq:beta_binom}), comparing to 1, 4, or 8 prior reference runs.
Reference runs must be labeled as good by PPD, and must have lasted at least 30 minutes.
This matches typical shifter procedures, where longer reference runs with no known issues are compared to the most recent data run.
The PCA and AE algorithms were trained on all 216 good runs from 2022 lasting at least 30 minutes, and the $\chi^{2~\prime}$ from Eq.~\ref{eq:mod_chi2}
is used as the anomaly score.
However, the AE failed to properly reconstruct certain classes of L1T histograms even in good runs, so the AE was excluded from the final global assessment.
The anomaly scores $s$ for each histogram from the 265 good runs (including those lasting less than 30 minutes) are ranked, allowing us to set variable anomaly thresholds $t_i = (s_i + s_{i+1})/2$ for each histogram type, based on the ranked values.
At the lowest threshold for a given histogram type, $t_0 = (s_0 + s_1)/2$, that histogram will be flagged as anomalous in all but one good run.
At $t_1 = (s_1 + s_2)/2$, this histogram will be flagged in all but two good runs, with the pattern continuing to higher thresholds.
The maximum threshold for a given histogram type is set \textit{above} the highest value seen in any good run: $t_{max} = s_{max} + (s_{max} - s_{max-1}) / 2$.
For each good and bad run, at each threshold index, we count how many of the 62 different histograms are flagged as anomalous, i.e.\ for which $s \geq t_i$.

Ideally AutoDQM would flag a large number of histograms from bad runs, and only a few from good runs.
It is also important to check what fraction of good and bad runs have a significant number of anomaly flags.
If good runs consistently have many histograms flagged as anomalous, ``alert fatigue'' will set in, and shifters will start to ignore the AutoDQM output.
Having a low number of flagged histograms is also important to allow shifters and experts to follow up on notable anomalies even in good runs, in case some intervention is needed.

\begin{figure}[htbp]
\begin{center}
\includegraphics[width=0.482\textwidth]{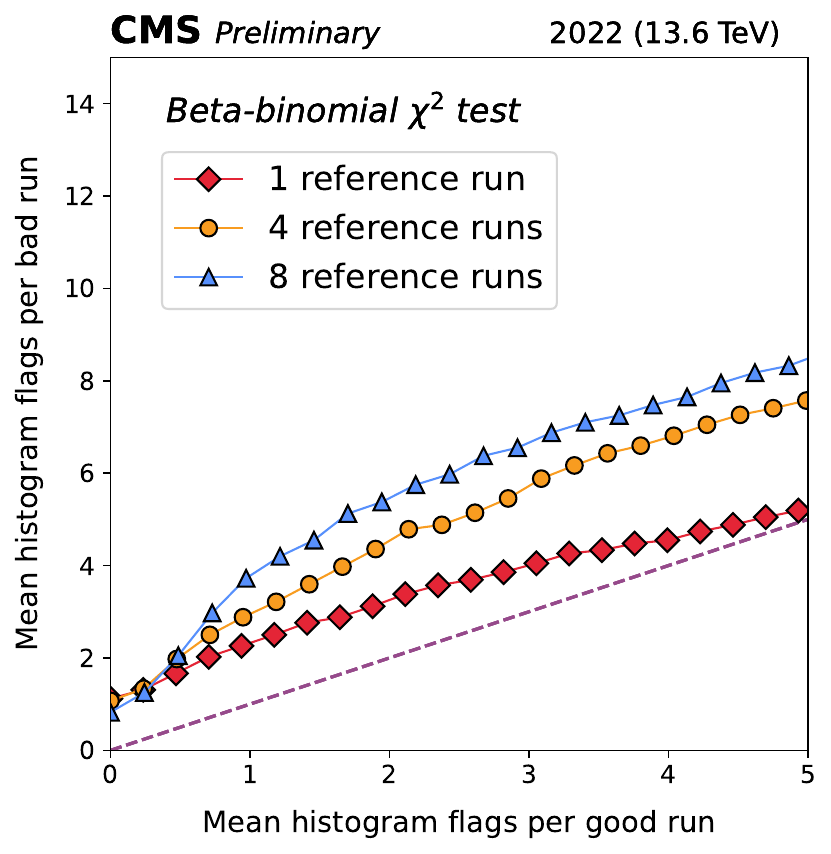}
\includegraphics[width=0.498\textwidth]{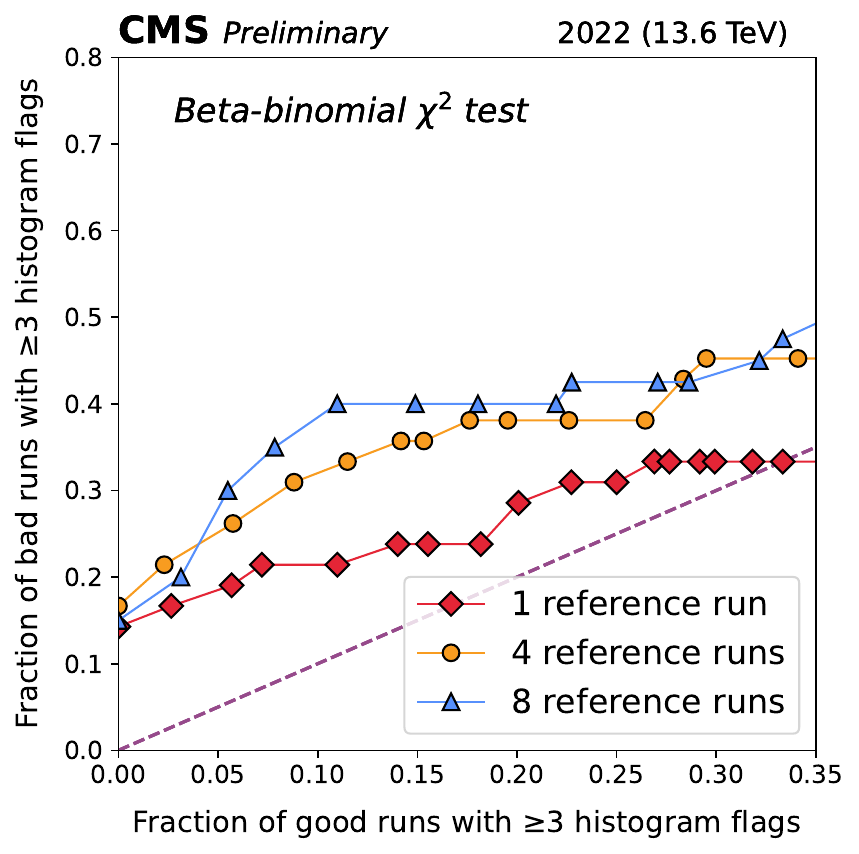}
\includegraphics[width=0.482\textwidth]{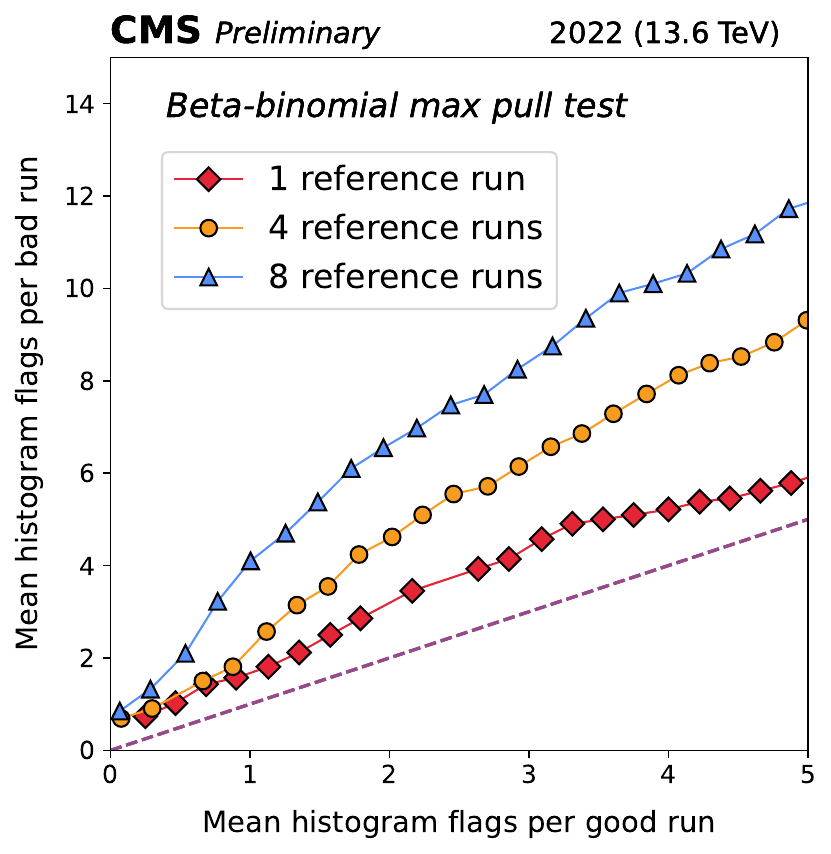}
\includegraphics[width=0.498\textwidth]{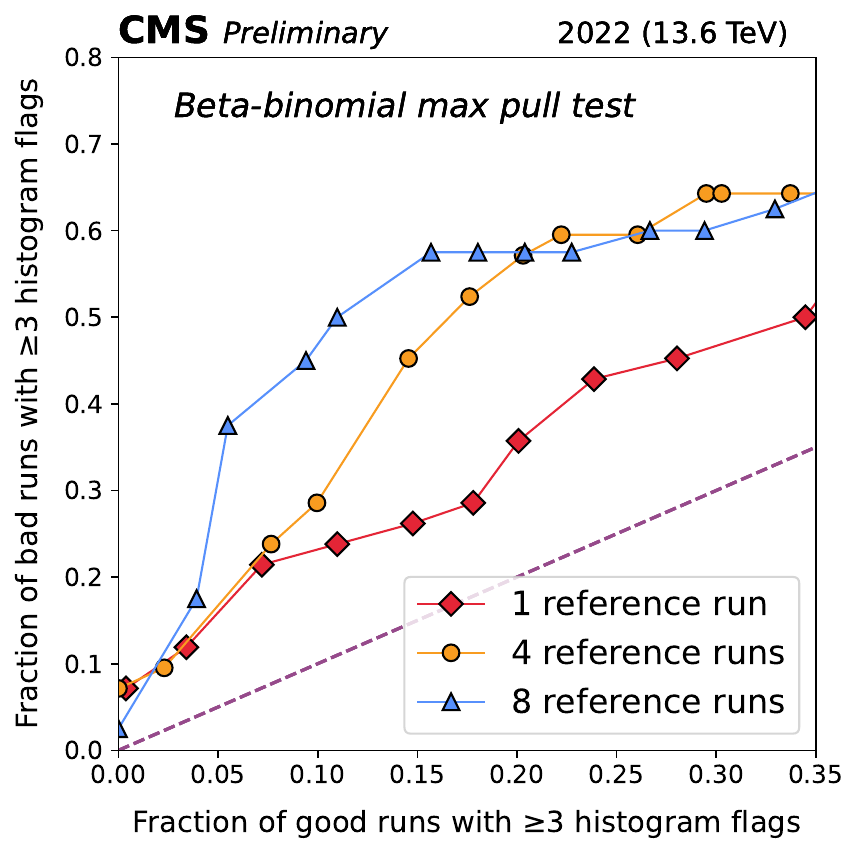}
\end{center}
\caption{Performance of the beta-binomial $\chi^2$ and maximum single-bin pull ($Z'_{max}$) statistical tests (Sec.~\ref{sec:autodqm_stat}) on L1T DQM histograms from 308 runs containing 2022 data.
ROC curves are constructed based on the mean number of $histograms$ flagged per run (left), and the fraction of $runs$ with at least 3 histograms flagged (right), comparing the data to 1, 4, or 8 prior reference runs.
}
\label{fig:beta_l1t_rocs}
\end{figure}

\begin{figure}[htbp]
\begin{center}
\includegraphics[width=0.482\textwidth]{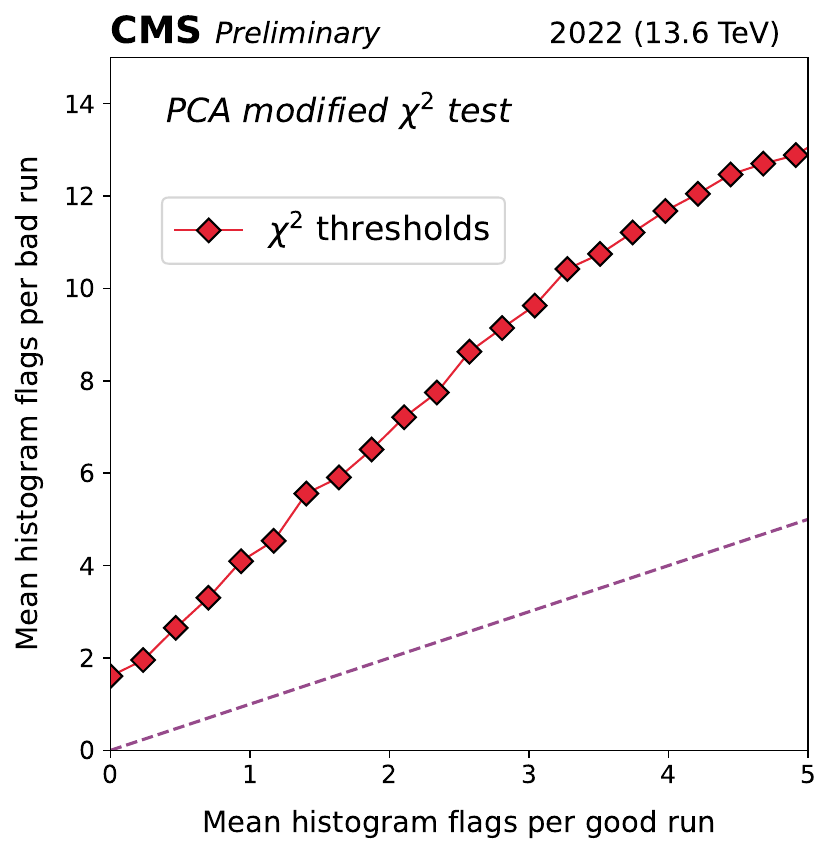}
\includegraphics[width=0.498\textwidth]{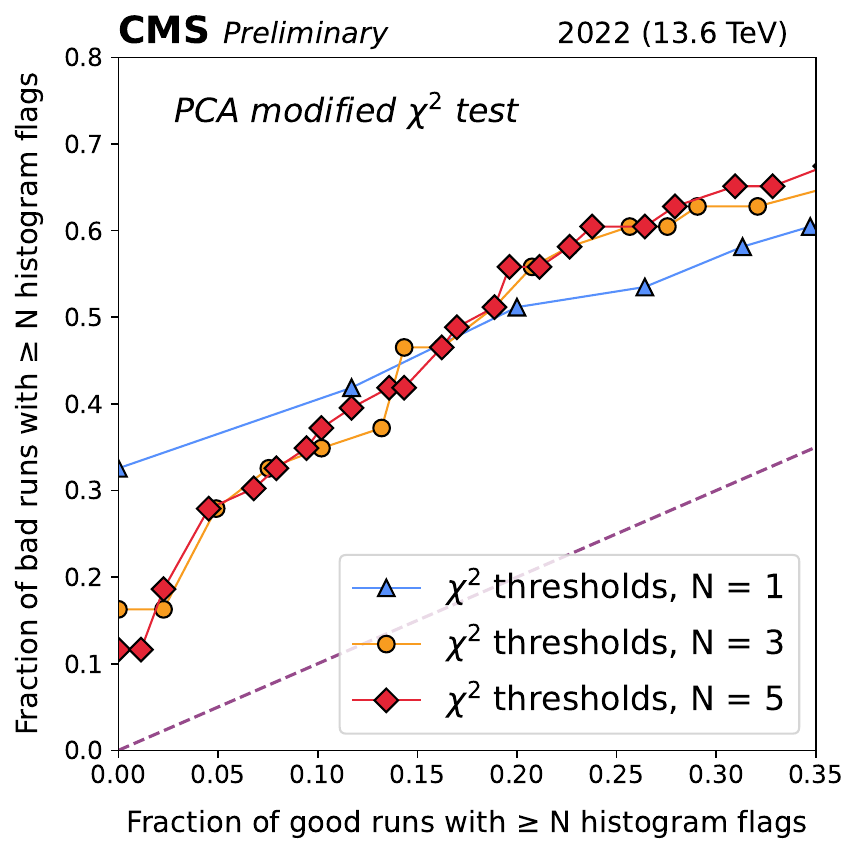}
\end{center}
\caption{Performance of the PCA modified $\chi^2$ test (Sec.~\ref{sec:autodqm_ml}) on L1T DQM histograms from 308 runs containing 2022 data.
ROC curves are constructed based on the mean number of $histograms$ flagged per run (left), and the fraction of $runs$ with at least 1, 3, or 5 histograms flagged (right).
}
\label{fig:pca_l1t_rocs}
\end{figure}

We thus construct two types of receiver operating characteristic (ROC) curves based on the number of flagged histograms in each run.
The histogram flags (HF) ROC shows the average number of $histograms$ flagged as anomalous by a given test for a variety of thresholds in good and bad runs, as shown on the left-hand side of Figs.~\ref{fig:beta_l1t_rocs} -- \ref{fig:pca_l1t_rocs}.
The run flags (RF) ROC is based on the fraction of $runs$ with at least $N$ histograms (out of the 62) flagged as anomalous, and can be seen on the right-hand side of Figs.~\ref{fig:beta_l1t_rocs} -- \ref{fig:pca_l1t_rocs}.
We evaluate the RF ROC for $N =$ 1, 3, and 5, to test both ``tight'' and ``loose'' anomaly thresholds.

\subsection{Assessment results}
\label{sec:metastudy_results}

The beta-binomial and PCA algorithms both show strong discrimination between good and bad runs.
When the mean number of histogram flags is low ($< 3$) for good runs, the HF ROC plots show 3 -- 4 times more flags in bad runs.
The RF ROC plots show similar behavior for the fraction of good and bad runs with $N \geq$ 1, 3, and 5 histogram flags: with less than 12\% of good runs flagged, 35 -- 50\% of bad runs exceed the threshold.
It is worth noting that we do not expect AutoDQM (or any anomaly detection approach) to identify 100\% of bad runs in this data set.
In many cases, the relevant issue had no impact on the L1T, or was simply not visible in the online DQM histograms.
We also do not expect to achieve a 0\% flagging rate for good runs.
In fact, many good runs have true anomalies which $should$ be flagged.
Nevertheless, AutoDQM applied to L1T monitoring alone was able to detect half of the serious issues affecting CMS data quality in 2022, with less than 12\% of good runs flagged as anomalous.

The beta-binomial $\chi^2$ and $Z'_{max}$ tests perform significantly better when using a larger number of reference runs for the comparison (Fig.~\ref{fig:beta_l1t_rocs}).
This is expected, as reconstructed particle occupancy distributions in the detector are sensitive to the number of simultaneous collisions (``pileup''), so runs taken with different amounts of pileup yield DQM histograms with different shapes.
When using multiple reference runs, at least one of them is likely to have similar pileup conditions to the data run being tested.
The pileup dependence is taken into account naturally in the AE and PCA algorithms (Fig.~\ref{fig:pca_l1t_rocs}), which are trained on a large number of runs spanning the full range of pileup conditions.

\begin{figure}[htbp]
\begin{center}
\includegraphics[width=0.482\textwidth]{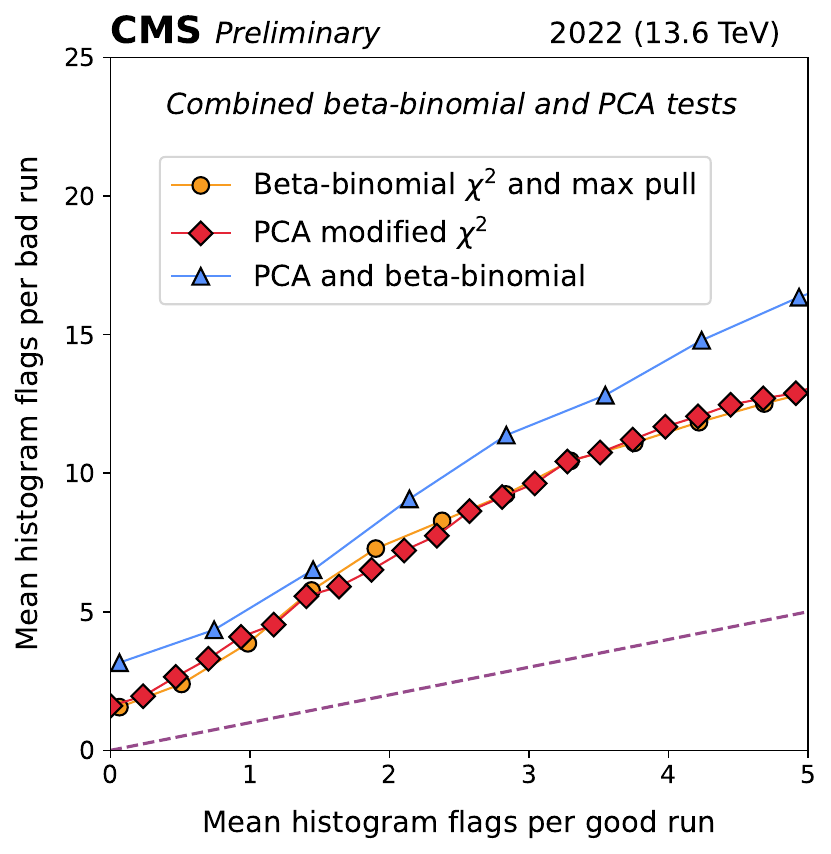}
\includegraphics[width=0.498\textwidth]{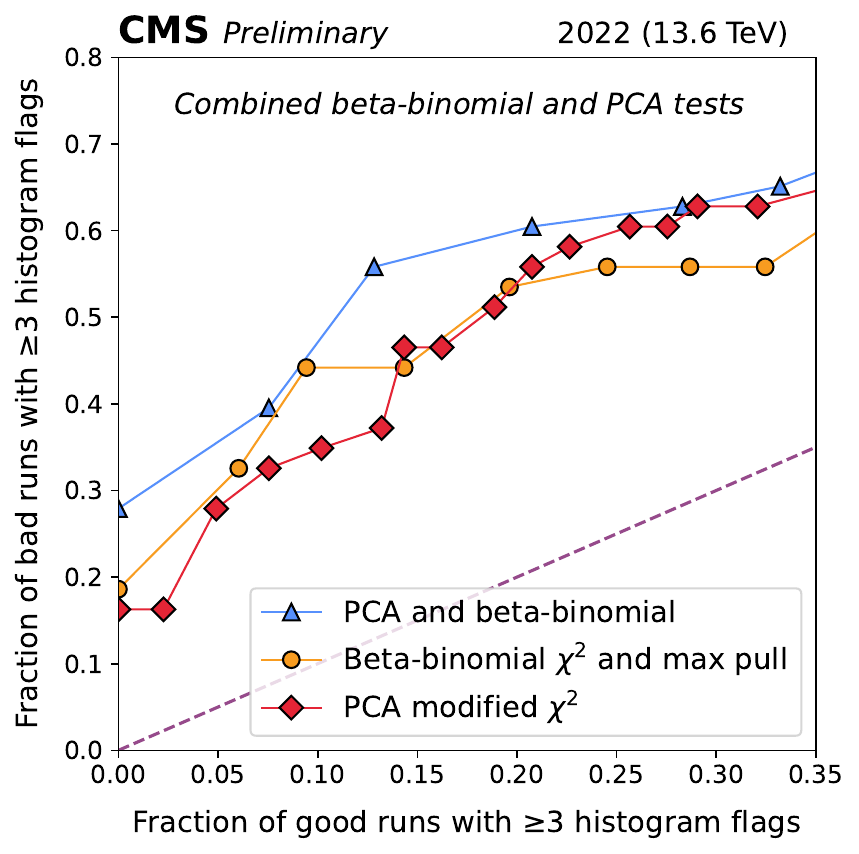}
\end{center}
\caption{Performance of the combined beta-binomial and PCA algorithms on L1T DQM histograms from 308 runs containing 2022 data.
ROC curves are constructed based on the mean number of $histograms$ flagged per run (left), and the fraction of $runs$ with at least 3 histograms flagged (right).
}
\label{fig:combined_algos}
\end{figure}

While there is some variation among the algorithms, none appears to be decisively superior to the others.
The best performance is achieved by applying all three quality tests simultaneously (Fig.~\ref{fig:combined_algos}).
In this case, the HF ROC plot shows 4 -- 6 times more flags in bad runs than in good runs, and over 55\% of bad runs have at least 3 flags for a threshold where only 13\% of good runs have 3 flags.
For the combined tests, the number of distinct $flags$ is counted, i.e.\ if the same histogram is flagged by 2 tests, that counts as 2 anomaly flags.


\subsection{Application to muon detector monitoring}
\label{sec:muons}

\begin{figure}[htbp]
\begin{center}
\includegraphics[width=0.75\textwidth,angle=-90]{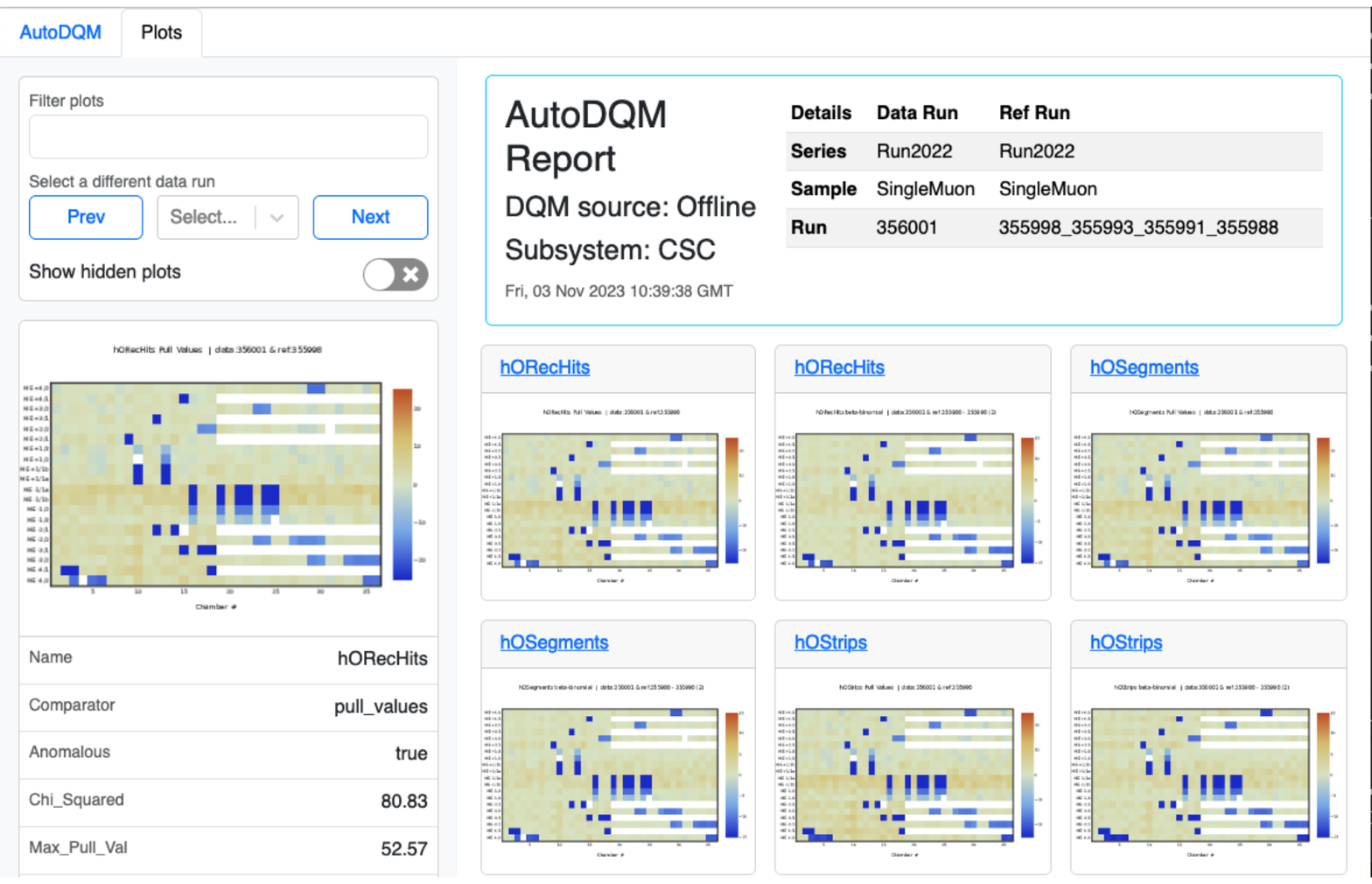}
\end{center}
\caption{AutoDQM GUI webpage for the cathode strip chambers (CSCs) in run 356001 in 2022, showing numerous chambers with anomalously low occupancy of reconstructed muon ``hits'' in blue.
Each plot can be expanded by clicking, and histograms not flagged as anomalous can be viewed using the ``Show hidden plots'' toggle.
The precise anomaly scores for each histogram are displayed in a panel on the left. }
\label{fig:CSC_webpage}
\end{figure}

\begin{figure}[htbp]
\begin{center}
\includegraphics[width=0.49\textwidth]{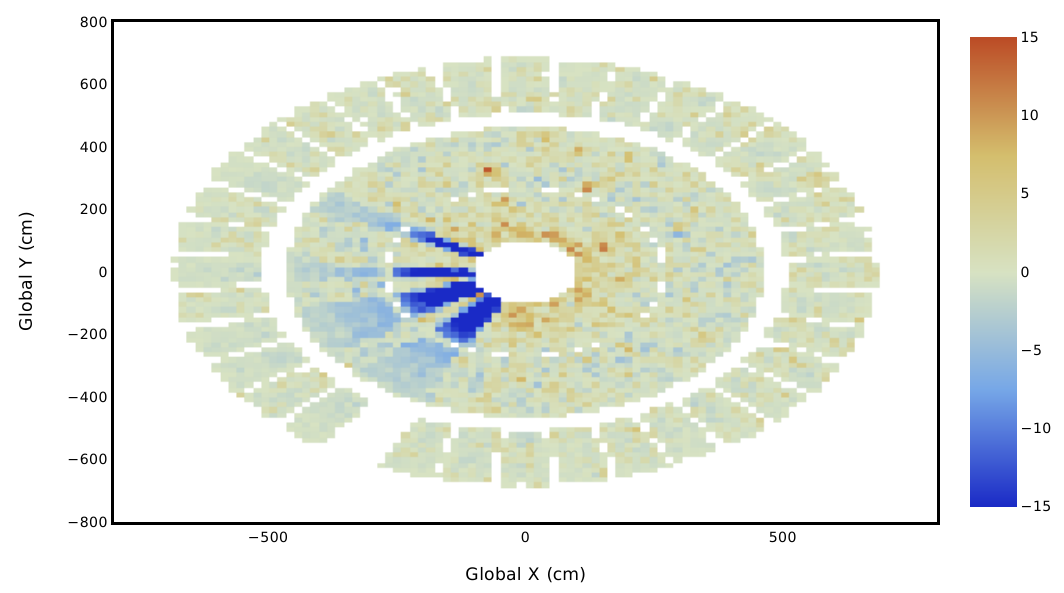}
\includegraphics[width=0.49\textwidth]{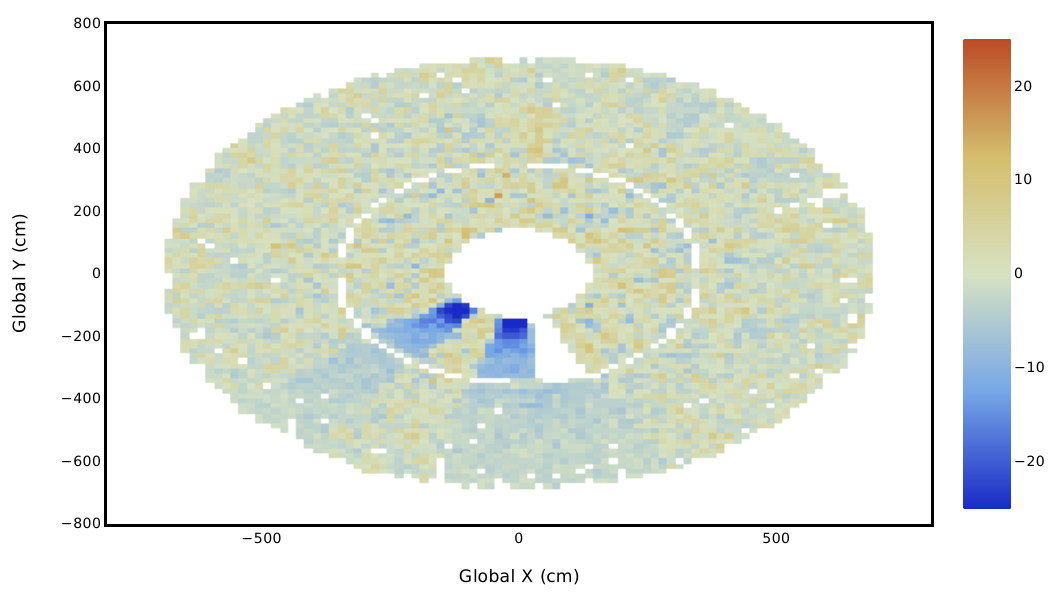}
\end{center}
\caption{AutoDQM GUI plots of the geometrical reconstructed muon ``hit'' distribution in the CSC detectors for run 356001 in 2022, showing regions with anomalously low occupancy in blue.
Regions which are consistently empty across multiple runs appear in white, allowing the shifter to distinguish between new and long-running issues. }
\label{fig:CSC_RecHit}
\end{figure}

While the L1T online DQM histograms provide a good test case for global AutoDQM performance with inputs from multiple subdetectors, the AutoDQM tool has also been applied successfully in muon detector monitoring.
Of the 43 bad runs used in the studies above, only 1 was flagged as bad by PPD due to issues in the muon detectors.
Nevertheless, it is important to closely monitor and identify any significant changes in the muon detector performance, e.g.\ chambers which occasionally malfunction, requiring expert intervention.
The CSCs in the CMS endcaps contain a total of 540 chambers, of which a handful may be disabled at any given time.
Very rarely, a dozen or more chambers temporarily malfunction simultaneously.
In this case, the AutoDQM webpage flags numerous CSC DQM histograms as anomalous (Fig.~\ref{fig:CSC_webpage}).
Furthermore, the individual AutoDQM plots clearly show the geometrical regions with new deficits of muon tracks in blue (Fig.~\ref{fig:CSC_RecHit}).
This allows CSC experts to quickly assess the scope and identify the source of new detector issues, enabling prompt intervention.

\section{Summary and outlook}
\label{sec:summary}

Data quality monitoring (DQM) presents an immense challenge to particle physics experiments, which will only grow as the data collected increases in volume and complexity.
The AutoDQM system for generalized, automated DQM dramatically augments the ability of physicists to quickly identify and localize anomalous behavior in the CMS detector.
Using a set of monitoring histograms from the CMS Level-1 Trigger system covering the entire CMS proton-proton collision data set from 2022, AutoDQM's combined statistical and machine learning tests successfully identified over 50\% of all ``bad'' data with significant detector malfunction, while flagging less than 15\% of ``good'' data as anomalous.
AutoDQM also demonstrates its effectiveness in visually highlighting changes in CMS muon detector performance.
Application to additional CMS subdetector systems will allow for more rapid, accurate identification of important issues affecting collision data in the future.

\bibliographystyle{JHEP}

\newpage
\bibliography{references.bib}

\newpage
\appendix

\section{Anomaly score dependence on histogram occupancy}
\label{sec:appendix}
\begin{figure}[htbp]
\begin{center}
\includegraphics[width=0.33\textwidth]{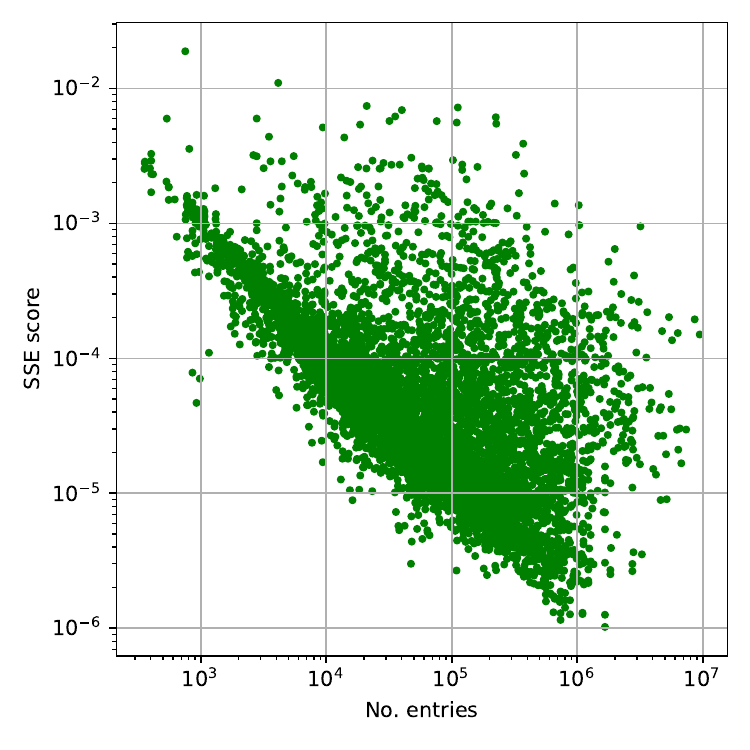}
\includegraphics[width=0.33\textwidth]{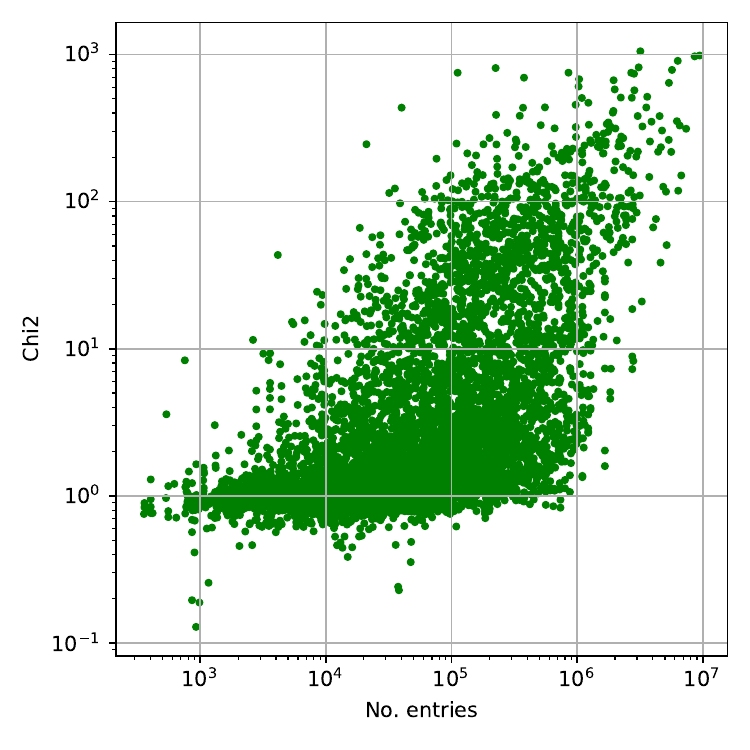}
\includegraphics[width=0.33\textwidth]{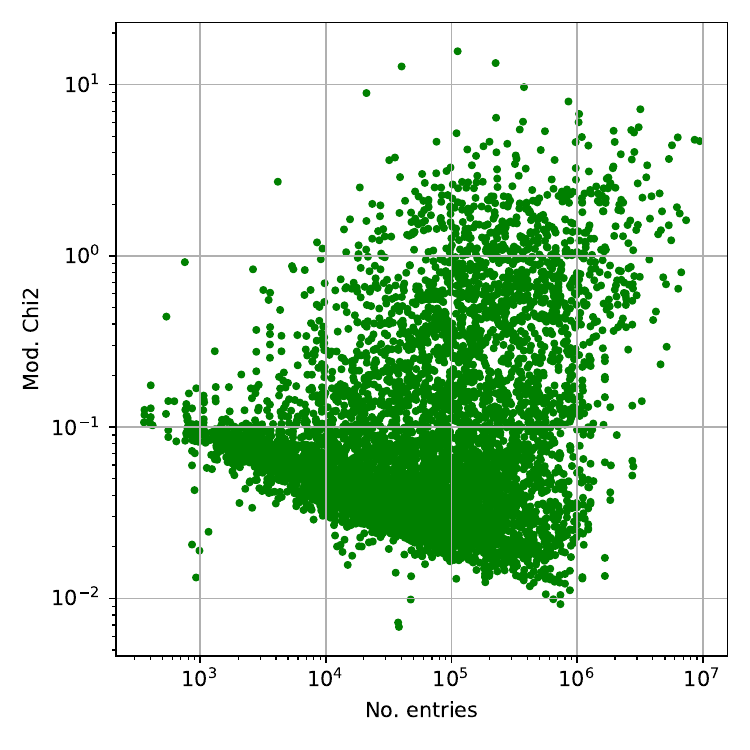}
\includegraphics[width=0.33\textwidth]{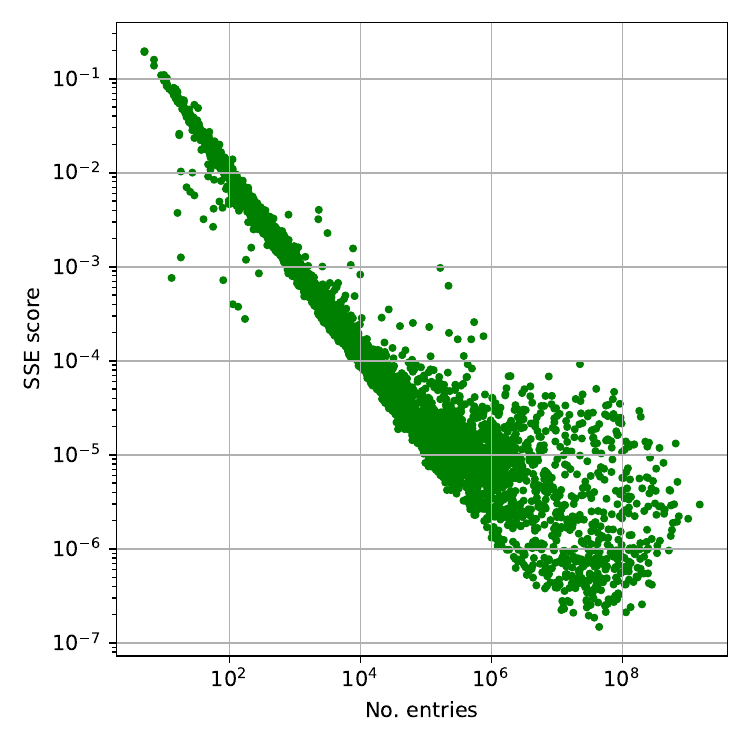}
\includegraphics[width=0.33\textwidth]{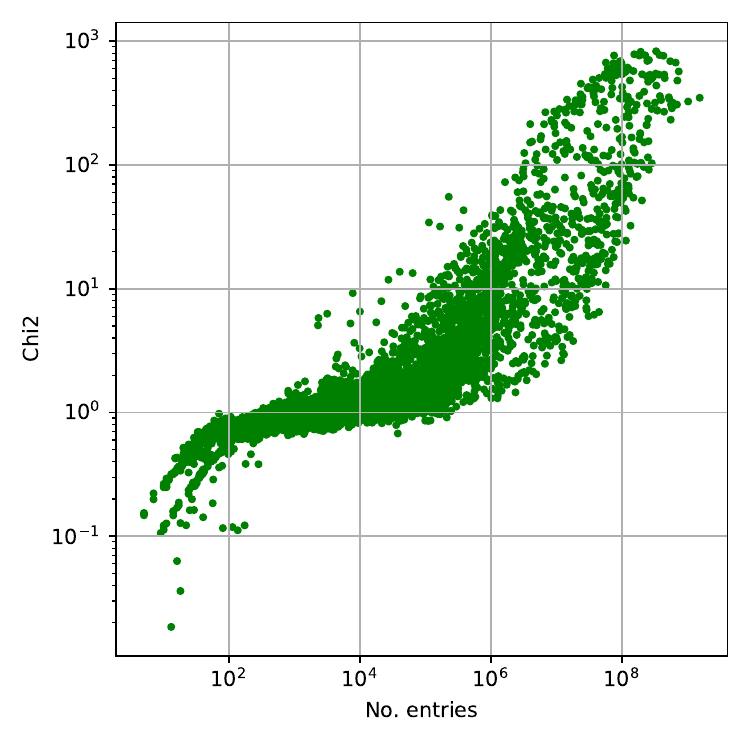}
\includegraphics[width=0.33\textwidth]{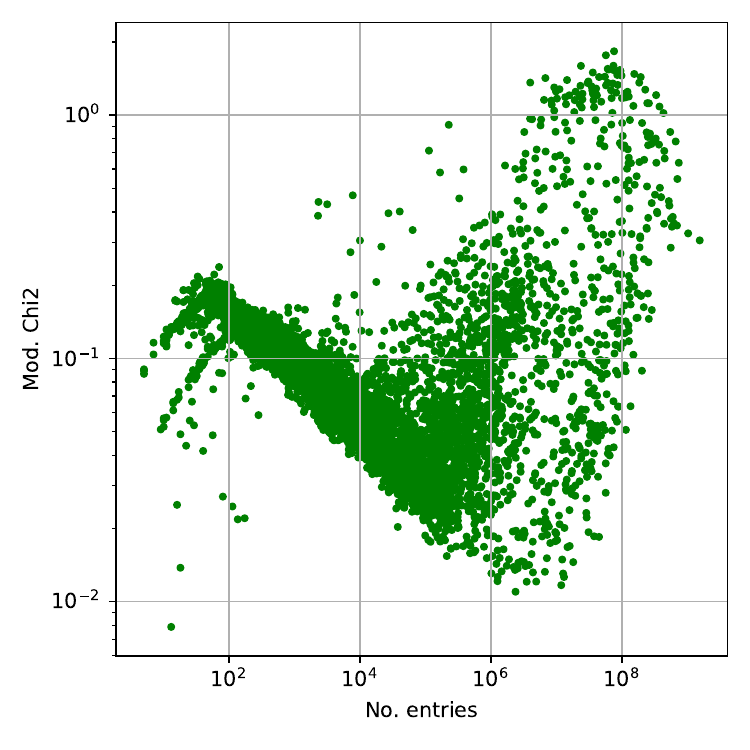}
\end{center}
\caption{2D correlation of PCA derived SSE (left), $\chi^2$ (middle), and $\chi^{2~\prime}$ (right) score values to the number of input data histogram entries for 1D (top) and 2D (bottom) L1T DQM histograms from good runs collected by CMS in 2022.
The $\chi^{2~\prime}$ test has the smallest correlation to the number of entries, and the range of $\chi^{2~\prime}$ values is orders of magnitude smaller than the SSE and $\chi^2$ ranges.
}
\label{fig:PCA_2D_score}
\end{figure}

\begin{figure}[htbp]
\begin{center}
\includegraphics[width=0.34\textwidth]{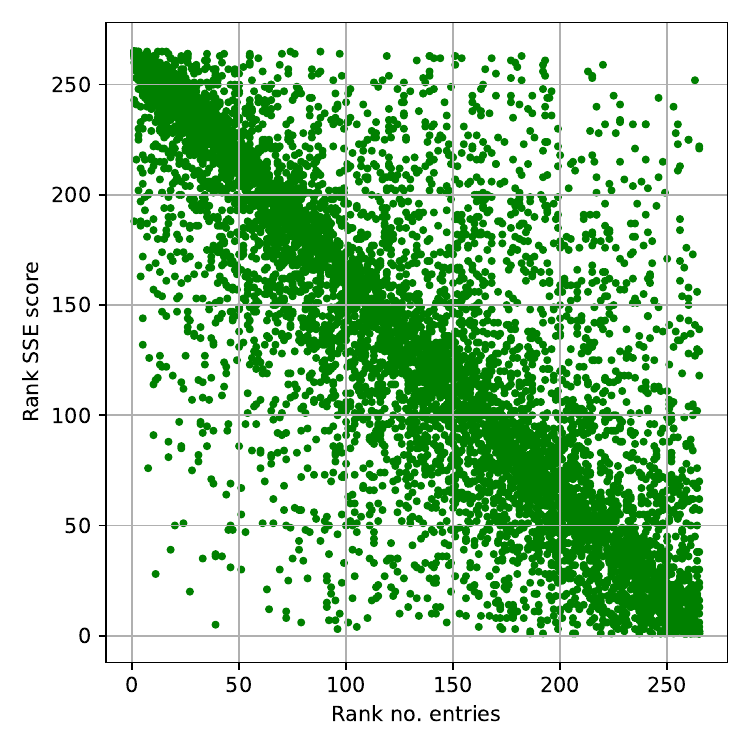}
\includegraphics[width=0.34\textwidth]{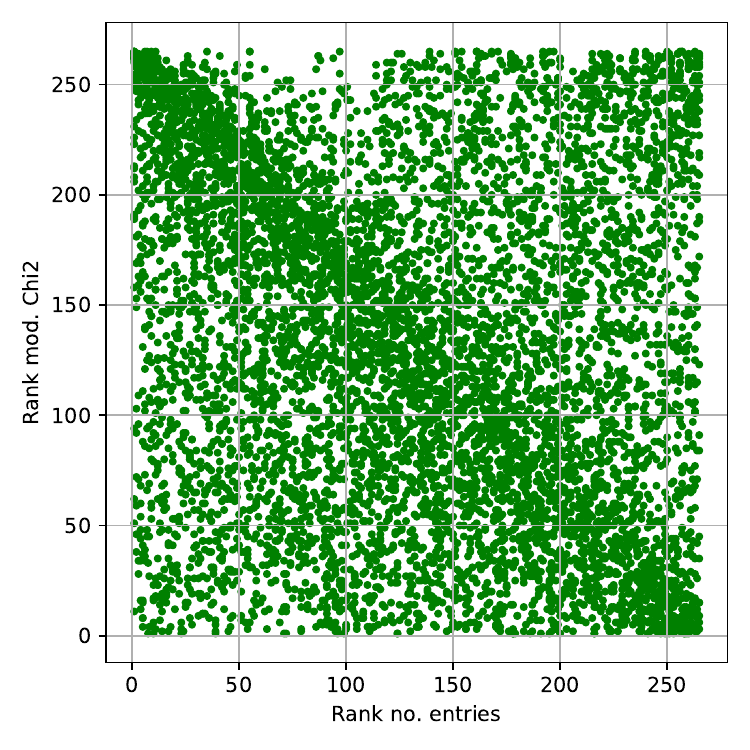}
\includegraphics[width=0.34\textwidth]{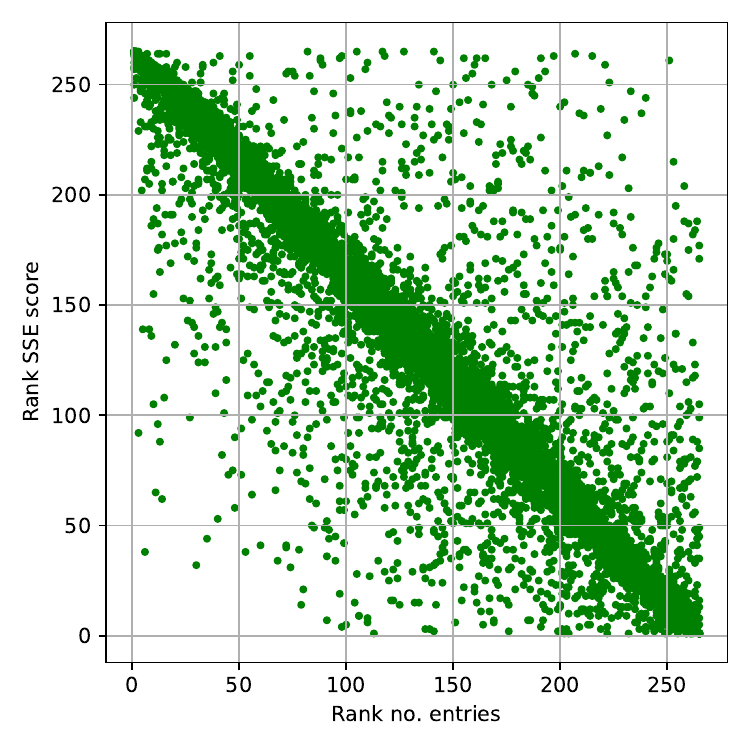}
\includegraphics[width=0.34\textwidth]{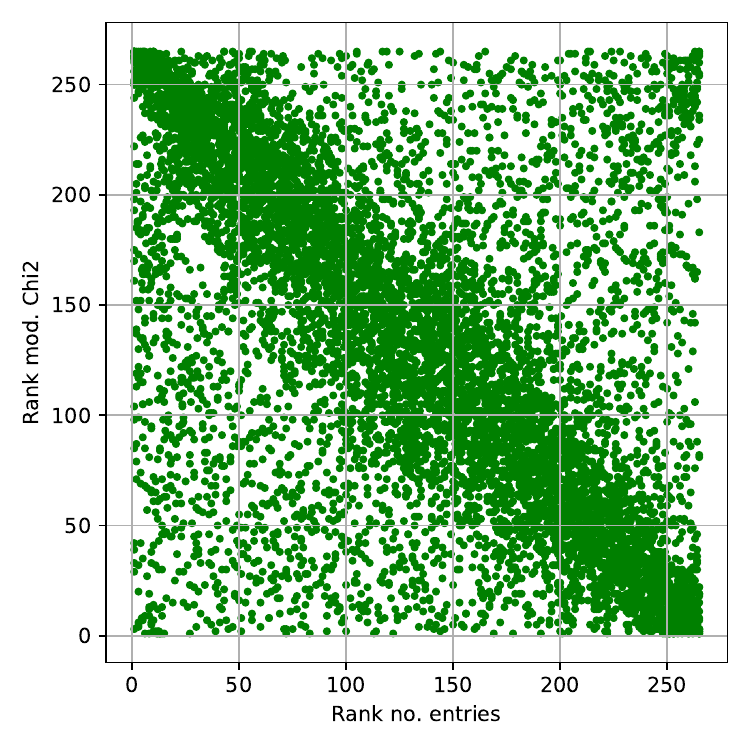}
\end{center}
\caption{2D correlation of PCA derived SSE (left) and $\chi^{2~\prime}$ (right) score ranks to the rank of the number of input data histogram entries for 1D (top) and 2D (bottom) L1T DQM plots from 265 good runs collected by CMS in 2022.
The rank is computed by comparing each histogram to histograms of the same type from other good runs, with 62 histogram types in total.
The $\chi^{2~\prime}$ test has much lower correlation to the number of entries than SSE, and high-rank scores appear for histograms with both high- and low-rank number of entries.
}
\label{fig:PCA_2D_rank}
\end{figure}

In order effectively monitor data quality, anomaly tests should be unbiased with respect to the number of entries in the histogram, which correlates directly to the length of the data-taking run and the collision rate.
As shown in Fig.~\ref{fig:PCA_2D_score} (left), the SSE values (Eq.~\ref{eq:sse}) comparing the normalized input histograms and the PCA transformed output histograms in good runs show a strong linear anti-correlation with the number of entries in the original data histograms.
This means that histograms with lower statistics (from shorter runs) will have a much higher probability to be flagged as anomalous.
Comparing the unnormalized input data histogram to a high-statistics scaling of the PCA transformed output histogram using a beta-binomial $\chi^2$ test (Eqs.~\ref{eq:beta_binom} and \ref{eq:beta_binom_PCA}), the trend is opposite: only high-statistics histograms get very high $\chi^2$ scores (Fig.~\ref{fig:PCA_2D_score} middle).
After testing numerous approaches, the modified $\chi^{2~\prime}$ value (Eq.~\ref{eq:mod_chi2}), which is scaled by the number of entries to the $-1/3$ power, shows the least correlation with the number of entries (Fig.~\ref{fig:PCA_2D_score} right).
This is verified in Fig.~\ref{fig:PCA_2D_rank}, where the score rank within a given histogram class (similar to the threshold logic in Sec.~\ref{sec:metastudy_metrics}) is plotted against the rank of the number of entries.
Almost all of the high-rank SSE scores cluster at low-rank number of entries (left), while high-rank $\chi^{2~\prime}$ scores are found across the spectrum of ranked number of entries (right).

\begin{figure}[htbp]
\begin{center}
\includegraphics[width=0.74\textwidth]{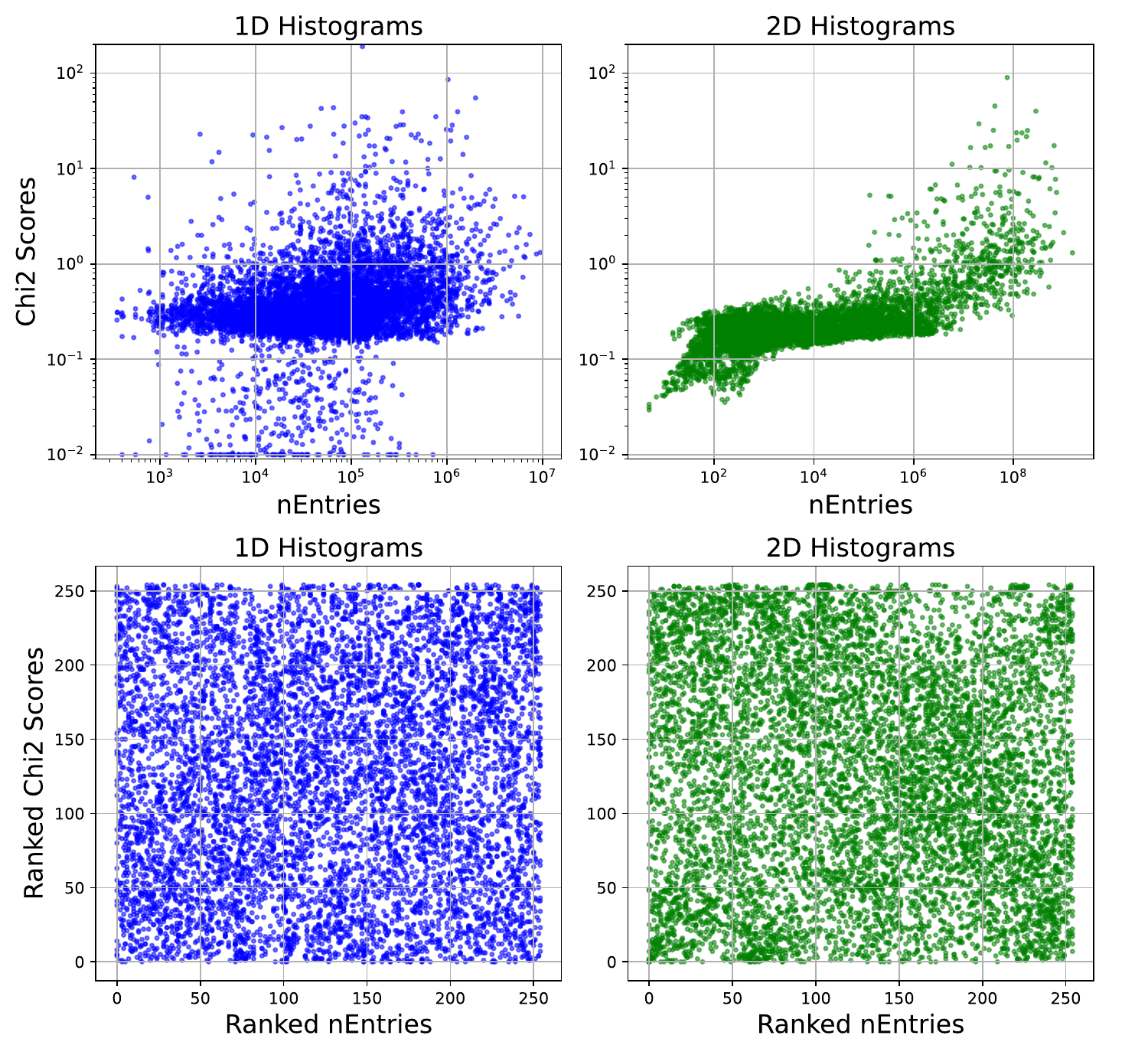}
\end{center}
\caption{2D correlation of beta-bimomial $\chi^2$ statistical test score values (top) and score ranks (bottom) to the number of input data histogram entries for 1D (left) and 2D (right) L1T DQM histograms from 265 good runs collected by CMS in 2022.
The rank is computed by comparing each histogram to histograms of the same type from other good runs, with 62 histogram types in total.
No strong correlation between rank and number of entries is observed.
}
\label{fig:BB_2D_chi2}
\end{figure}

\begin{figure}[htbp]
\begin{center}
\includegraphics[width=0.74\textwidth]{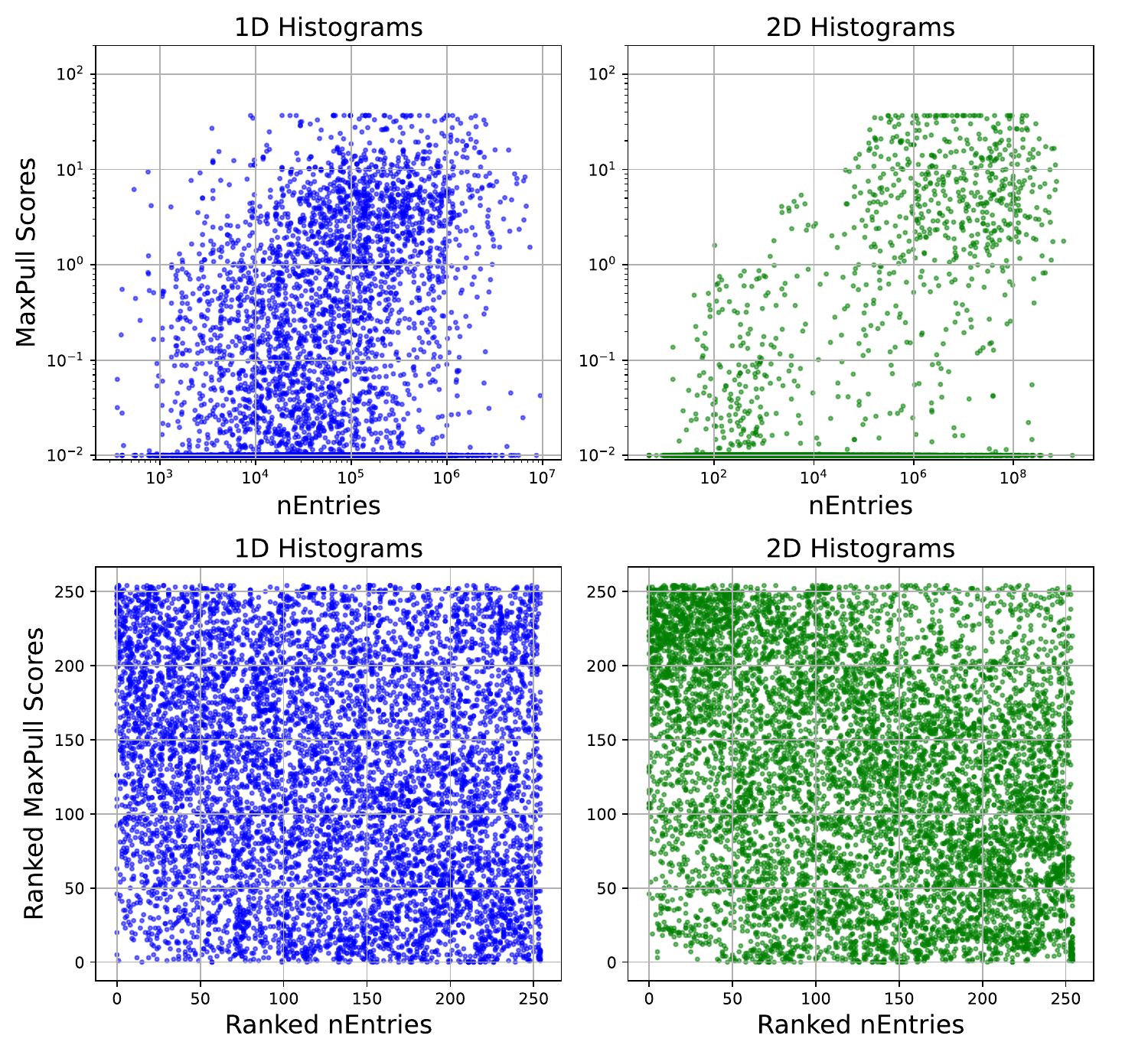}
\end{center}
\caption{2D correlation of beta-bimomial maximum single-bin pull ($Z'_{max}$) statistical test score values (top) and score ranks (bottom) to the number of input data histogram entries for 1D (left) and 2D (right) L1T DQM histograms from 265 good runs collected by CMS in 2022.
The rank is computed by comparing each histogram to histograms of the same type from other good runs, with 62 histogram types in total.
No strong correlation between rank and number of entries is observed.
}
\label{fig:BB_2D_pull}
\end{figure}

Figures~\ref{fig:BB_2D_chi2} and \ref{fig:BB_2D_pull} show that the beta-binomial $\chi^2$ and maximum single-bin pull ($Z'_{max}$ from Sec.~\ref{sec:autodqm_stat}) statistical tests (Eq.~\ref{eq:beta_binom}), performed with 8 reference runs, are not correlated to the number of entries in the data histogram being tested.

\end{document}